\documentclass[11pt,draftcls,onecolumn]{IEEEtran}

\usepackage[dvips]{epsfig}
\usepackage[dvips]{graphicx}
\usepackage{boxedminipage}
\usepackage{color}
\usepackage{amsfonts}
\usepackage{amssymb}
\usepackage{amsmath}
\usepackage{graphics,amsmath,amsfonts,amssymb,epsfig,latexsym,amsfonts,graphicx,amssymb}

\newcommand{\bfxi}{{\mbox{\boldmath $\xi$}}}
\newcommand{\bfepsilon}{{\mbox{\boldmath $\epsilon$}}}

\newcommand{\bfPhi}{{\mbox{\boldmath $\Phi$}}}

\newcommand{\bfp}{{\mbox{\boldmath $p$}}}

\newcommand{\bfUpsilon}{{\mbox{\boldmath $\Upsilon$}}}

\def\BibTeX{{\rm B\kern-.05em{\sc i\kern-.025em b}\kern-.08em
    T\kern-.1667em\lower.7ex\hbox{E}\kern-.125emX}}

\title{Averaged Iterative Water-Filling Algorithm: Robustness and Convergence}
\author{Mingyi Hong, Alfredo Garcia
\thanks{M. Hong and A. Garcia are with the Department of Systems and Information Engineering,
University of Virginia, Charlottesville, VA, 22903 USA (e-mail:
mh4tk@virginia.edu, ag7s@virginia.edu)}}
\begin{document}
\maketitle
\begin{abstract}
The convergence properties of the Iterative water-filling (IWF)
based algorithms (\cite{yu02}, \cite{scutari08a}, \cite{wang08})
have been derived in the {ideal situation} where the transmitters in
the network are able to obtain the \emph{exact} value of the
interference plus noise (IPN) experienced at the corresponding
receivers \emph{in each iteration} of the algorithm. However, these
algorithms are not robust because they diverge when there is {\it
time-varying} estimation error of the IPN, a situation that arises
in real communication system. In this correspondence, we propose an
algorithm that possesses convergence guarantees in the presence of
various forms of such time-varying error. Moreover, we also show by
simulation that in scenarios where the interference is strong, the
conventional IWF diverges while our proposed algorithm still
converges.
\end{abstract}

\section{Introduction}
\subsection{The IWF Algorithm}
The Iterative Water-Filling algorithm has been first proposed by Yu
et al in \cite{yu02} to solve the power allocation problem in DSL
network, and it has since been applied to various areas in
communications and signal processing to obtain solutions for network
power allocation problems (see, e.g. \cite{wang08},
\cite{scutari08b}, \cite{scutari08c}, \cite{luo06b} and the
references therein).

We consider an application of the IWF algorithm to the resource
allocation problem in wireless communication network, where there
are $N$ users and $K$ subchannels; each user is a
transmitter-receiver pair that tries to communicate with each other.
Define the sets $\mathcal{N}\triangleq\{1,\cdots,N\}$, and
$\mathcal{K}\triangleq\{1,\cdots,K\}$; let
$\{S_i\}_{i\in\mathcal{N}}$ denote the set of users in the network;
let $p_i(k)$ denote the amount of power $S_i$ transmits on channel
$k$; let $\mathbf{p}_i\triangleq [p_i(1),\cdots,p_i(K)]^\intercal$,
$\mathbf{p}_{-i}\triangleq
\left[\mathbf{p}_1^\intercal,\cdots,\mathbf{p}_{i-1}^\intercal,\mathbf{p}_{i+1}^\intercal,
\cdots,\mathbf{p}_{N}^\intercal\right]^{\intercal}$ and
$\mathbf{p}\triangleq\left[\mathbf{p}_1^\intercal,\cdots,\mathbf{p}_N^\intercal\right]^{\intercal}$.
The channel gain between the transmitter of $S_i$ to the receiver of
$S_j$ on channel $k$ is denoted by $|H_{i,j}(k)|^2$. The power of
the environmental noise experienced at $S_i$'s receiver on channel
$k$ is denoted by $n_i(k)$. We assume that there is no interference
cancelation performed at the receivers, and the interference caused
by the other users is considered as noise. Then the signal to
interference plus noise ratio (SINR) measured at the receiver of
$S_i$ on channel $k$ can be expressed as: $
SINR_i(k)=\frac{|H_{i,i}(k)|^2 p_i(k)}{n_i(k)+\sum_{j\ne i}
|H_{j,i}(k)|^2 p_j(k)}.$

Using Shannon's capacity, the maximum transmission rate achievable
for $S_i$ can be expressed as:
$R_i(\mathbf{p}_{i},\mathbf{p}_{-i})=\sum_{i=1}^{K}
\log(1+SINR_i(k)).$ We consider the following constraints for each
user: {\bf {[C-1)]}} each $S_i$ has limited power budget, i.e., $0
\leq \sum_{k=1}^{K}p_i(k)\leq \bar{p}_i,~\forall~i\in \mathcal{N}$;
{\bf [C-2)]} we require $0\leq p_i(k)\leq p_{mask}(k),
\forall~k\in\mathcal{K}$ and $i\in\mathcal{N}$. As such, we use
$\mathcal{P}_i$ to denote the set of feasible power allocations for
$S_i$: $ \mathcal{P}_i\triangleq
\left\{\mathbf{p}_i:\sum_{k=1}^{K}p_i(k)\leq \bar{p}_i, 0\le
p_i(k)\le p_{mask}(k),~\forall~ k\in \mathcal{K}\right\}$.

Dynamic power allocation in this network can be formulated as a
non-cooperative game where each user $S_i$ is interested in
maximizing its own rate when deciding how to allocate its power
across the spectrum, i.e., $S_i$ wants to find
$\mathbf{p}_i^*\in\mathcal{P}_i$ such that {  $
\mathbf{p}_i^*\in\arg\max_{\mathbf{p}_i\in\mathcal{P}_i}
{R}_i(\mathbf{p}_i,\mathbf{p}_{-i}). $}
%The problem of how to dynamically and distributedly find a suitable
%power allocation for the users across all the available bandwidth
%can be formulated into a noncooperative game, in which
%$\{S_i\}_{i\in\mathcal{N}}$ are the players, and they are interested
%in allocating their available power across the spectrum to {\it
%maximize their own rate}, i.e., $S_i$ wants to find
%$\mathbf{p}_i^*\in\mathcal{P}_i$ such that {  $
%\mathbf{p}_i^*\in\arg\max_{\mathbf{p}_i\in\mathcal{P}_i}
%{R}_i(\mathbf{p}_i,\mathbf{p}_{-i}). $}
A Nash Equilibrium (NE) can be expressed as the set of power
profiles $\{\mathbf{p}_i^{\ast}\}_{i\in\mathcal{N}}$ satisfying the
set of conditions: {  $
\mathbf{p}^{\ast}_i\in\arg\max_{\mathbf{p}_i\in\mathcal{P}_i}R_i(\mathbf{p}_i,\mathbf{p}_{-i}^{\ast})
~\forall~i\in\mathcal{N}. $}  The IWF and its various extensions are
essentially policies for the players to jointly reach a NE of this
game in a distributed manner.

In the IWF, the transmitters iteratively adjust their transmission
power levels to maximize their own transmission rate. Specifically,
in iteration $t+1$, each user $S_i$ computes {
$\left\{p^{t+1}_i(k)\right\}_{k\in\mathcal{K}}$} as follows:
\begin{align}
p^{t+1}_i(k)&=\arg\max_{\mathbf{p}_i\in\mathcal{P}_i}R_i(\mathbf{p}_i,\mathbf{p}_{-i}^t)\nonumber\\
&=\Big[ \sigma_i-\big(\bar{n}_i(k)+\sum_{j\ne
i}|\bar{H}_{j,i}(k)|^2p^t_j(k)\big)
\Big]^{p_{mask}(k)}_0\nonumber\\
\vspace{-0.2cm}&=\left[ \sigma_i-IPN^t_i(k))
\right]^{p_{mask}(k)}_0\triangleq
\Phi^k_i(\mathbf{p}^t_{-i})\label{eqWaterFilling}
\end{align}
where $\sigma_i$ is the dual variable associated with the total
power constraint for user $i$, and  it is also referred to as the
``water level" in the traditional water-filling algorithm;
$|\bar{H}_{j,i}(k)|^2$ and $\bar{n}_i(k)$ are defined as
$|\bar{H}_{j,i}(k)|^2\triangleq\frac{|H_{j,i}(k)|^2}{|H_{i,i}(k)|^2}$
and  $\bar{n}_i(k)\triangleq \frac{n_i(k)}{|H_{i,i}(k)|^2}$,
respectively; $IPN^t_i(k)$ is defined as the normalized total
interference plus noise (IPN) for user $S_i$ on channel $k$ at time
$t$: {  $IPN^t_i(k)\triangleq\bar{n}_i(k)+\sum_{j\ne
i}|\bar{H}_{j,i}(k)|^2p^t_j(k)$}. This quantity is measured at the
receivers and fed back to their corresponding transmitters in each
iteration $t$ before $p^{t+1}_i(k)$ is computed. Define {
$\bfPhi_i(\mathbf{p}_{-i})\triangleq
[\Phi^1_i(\mathbf{p}_{-i}),\cdots,\Phi^K_i(\mathbf{p}_{-i})]^\intercal$},
and let {  $\bfPhi(\mathbf{p})\triangleq
[\bfPhi_1(\mathbf{p}_{-1}),\cdots,\bfPhi_N(\mathbf{p}_{-N})]^\intercal$}.
The function $\bfPhi(.)$ is called the {\it water-filling operator}
of the system, and the IWF algorithm can be written concisely: {
$\mathbf{p}^{t+1}=\bfPhi(\mathbf{p}^{t})$}. If the algorithm reaches
a power profile $\mathbf{p}^*$ such that
$\mathbf{p}^{*}=\bfPhi(\mathbf{p}^{*})$, we say the IWF converges.
%In this correspondence, we mainly consider a specific version of IWF
%and R-IWF where the users update their power profiles
%\emph{simultaneously}. Extension of our analysis to other versions
%of IWF can be made similarly.

%The ``relaxed" version of IWF (R-IWF) mentioned in the introduction
%section is of the following form:
%\begin{eqnarray}
%\mathbf{p}^{t+1}=(1-\lambda)\mathbf{p}^t+\lambda{\bfPhi}(\mathbf{p}^{t})
%\end{eqnarray}
%where the $\lambda\in(0,1]$ is a free parameter. This algorithm is
%specifically designed to deal with estimation error, and was proven
%to have similar convergence condition as IWF.

Sufficient conditions for convergence of the IWF algorithm and its
various extensions have been widely studied, for example, in
\cite{scutari08b}, \cite{luo06b}, \cite{shum07}. Essentially, if the
interference received (generated) at the receiver (transmitter) of
each user is weak enough compared with the desired signal, then the
IWF converges. When these conditions are not met, it is possible
that the IWF diverges \cite{leshem09}.

\subsection{The Uncertainty of IPN and the Water-Filling Operator}
One of the key assumptions of the IWF based algorithms is that the
receivers can \emph{always} get the exact values of the IPN on each
channel {in each iteration of the algorithm}, and fed back to the
transmitters. This assumption is not valid in real communication
systems because the power of the noise/interference experienced at
the receivers needs to be {\it estimated} in each iteration, thus is
subject to time-varying estimation errors \cite{benedict67},
\cite{pauluzzi00}. Therefore, in each iteration of the algorithm, we
can only obtain a noisy version of the true solution of
\eqref{eqWaterFilling}, referred to as the {\it noisy water-filling
solution}, as:
%and the uncertainties associated with these estimates might be
%time-varying.
\begin{align}
p^{t+1}_i(k)&=\left[ \widehat{{\sigma}}_i-\widehat{IPN}^t_i(k)
\right]^{p_{mask}(k)}_0\triangleq
\widehat{\Phi}_i^k(\mathbf{p}^t_{-i})\label{eqNoisyPhi}
\end{align}
where $\widehat{IPN}^t_i(k)$ is the noisy (estimated) IPN for user
$S_i$ on channel $k$. Note that the uncertainty of the IPN leads to
the inaccuracy of the dual variable, as now it should satisfy {
$\sum_{k=1}^{K}\hspace{-0.1cm}\left[
\widehat{\sigma}_i-\widehat{IPN}^t_i(k)
\right]^{p_{mask}(k)}_0\hspace{-0.6cm}=\bar{p}_i$}, and
$\sigma_i\ne\widehat{\sigma}_i$ in general.

There is little work in the literature that addresses the impact of
such time-varying uncertainty of the IPN on the performance of the
IWF algorithm. In \cite{wang08}, \cite{scutari08b}, a ``relaxed"
version of IWF (R-IWF) was proposed to heuristically deal with
inaccurate IPN levels. In each iteration, the transmission power
levels are computed as $
\mathbf{p}^{t+1}=(1-\lambda)\mathbf{p}^t+\lambda{\bfPhi}(\mathbf{p}^{t}),
$ where the $\lambda\in(0,1]$ is a free parameter. Although it has
been shown in \cite{scutari08b} that this algorithm converges under
similar conditions as the IWF in situations {\it without} IPN
uncertainty, the effect of this algorithm in the presence of IPN
inaccuracy is not clear, and as we will see later in the simulation
section, the performance of R-IWF depends strongly on the choice of
$\lambda$. In \cite{setoodeh09}, a robust version of IWF is proposed
to deal with errors related to changes in the number of users and
their mobility. The algorithm guarantees an acceptable level of
performance under worst case conditions (i.e., the maximum possible
error of the IPN). This algorithm trades performance in favor of
robustness, thus the equilibrium solution obtained is generally less
efficient than that of the original IWF. In our work, we are
concerned with reaching the equilibrium solution of the original IWF
in the presence of IPN uncertainty. In \cite{gohary09b}, the authors
provide a probabilistically robust IWF to deal with the quantization
errors of the IPN at the receiver of each user. In this algorithm,
users allocate their powers to maximize their total rate for a large
fraction of the error realization. However, a specific distribution
of the error process is assumed in the derivation of the algorithm,
and such statistical information is usually not available in
practice (as suggested in section V of \cite{setoodeh09}). A recent
work \cite{cheng10} proposes algorithms for system with finite-state
Markov channel in interference network. The channel itself is
modeled as time-varying in this work, and the objective is to track
the time-varying equilibria. In the present paper, uncertainty of
the channel is due to imperfect receiver estimation of the value of
IPN as opposed to changes in the state of the channel.

In this correspondence, we propose an extension of the IWF algorithm
that is robust in the presence of time-varying IPN uncertainties.
Specifically, we model the uncertainty regarding to the IPN as
time-varying added noises, and show that the proposed algorithm
converges with probability 1 under some conditions on the
channel gains and the noise process. We verify the above claim by
simulation, and demonstrate the advantage of the proposed algorithm
with respect to the original IWF and the R-IWF. Additionally, we
show by simulation that in some strong interference channels where
the conventional IWF algorithm diverges, our proposed algorithm
still converges. This last result indicates that the convergence
condition of our algorithm may be further relaxed.

%Moreover, because the users need to measure these quantities {\it
%repetitively}, we can view the uncertainties regarding these
%quantities as time-varying.
%We note here that the sequence of power profiles
%$\{\mathbf{p}^t\}_{t=1}^{\infty}$ generated by the IWF algorithm can
%be sensitive to estimation error, and might not be able to converge
%to the NE of the game. We show in the simulation section that this
%is indeed the case. Our aim is to provide an extension to the IWF
%algorithm that is more robust to estimation error.

%The contribution of this work is two fold: \emph{1)} it represents
%one of the first efforts to analyze and improve the robustness of
%the IWF algorithm; \emph{2)} it is also among the first efforts to
%extend the IWF algorithm to deal with the situations when the
%conventional IWF diverges.

This correspondence is organized as follows. In section
\ref{secProposedAlgorithm} we introduce the proposed algorithm and
provide convergence analysis. In section \ref{secSimulation} we
demonstrate the performance of the proposed algorithm and compare
the results with conventional IWF. This correspondence concludes in
section \ref{conclusion}.

\section{Proposed Algorithm and Convergence
Results}\label{secProposedAlgorithm}
%In this section, we detail the
%propose algorithm that has nice convergence properties and is robust
%against uncertainties of the values of IPN.

In the proposed algorithm, in each iteration $t$, all the users
compute their power allocations as follows:
\begin{enumerate}
\item Obtain $\{\widehat{IPN}^t_i(k)\}_{k\in\mathcal{K}}$, and calculate the noisy water-filling solution
$\widehat{\Phi}_i(\mathbf{p}^t_{-i})$.

\item Calculate the power output according to the following policy: {
\begin{align}
\mathbf{p}^{t+1}_i= \left\{ \begin{array}{ll}
\widehat{\bfPhi}_i(\mathbf{p}^t_{-i}) &\textrm{for}~t= 0\\
(1-\alpha_t)\mathbf{p}^t_i+\alpha_t
\widehat{\bfPhi}_i(\mathbf{p}^t_{-i})  &\textrm{for}~t\ge 1 \\
\end{array} \right. \label{eqAIWF}
\end{align}}
\end{enumerate}

where the elements of $\widehat{\bfPhi}_i(\mathbf{p}^t_{-i})$ are
defined in \eqref{eqNoisyPhi}. The sequence
$\{\alpha_t:0<\alpha_t\le1\}_{t=0}^{\infty}$ satisfies the following
(define $\alpha_0=1$):
\begin{align}
\lim_{T\to\infty}\sum_{t=0}^{T}\alpha_t=\infty,~\lim_{T\to\infty}\sum_{t=0}^{T}\alpha_t^2<\infty\label{eqAlphaProperty3}.
\end{align}
%A typical choice of $\alpha_t$ is $\alpha_t=\frac{1+b}{t+c}$, with
%$b$, $c$ two positive constants, and $b\le c$.
Note that from the last inequality in \eqref{eqAlphaProperty3}, we
have $\lim_{t\to\infty}\alpha_t=0$. The update procedure
\eqref{eqAIWF} is essentially Mann's iterations (see \cite{mann53}
for its properties), which is designed for situations where
conventional iterative methods for finding the fixed point of a
self-mapping (say Picard's method) fail. If we choose
$\alpha_t=\frac{1}{t+1}$, then the update policy in \eqref{eqAIWF}
can be rewritten as: $ \mathbf{p}^{T+1}_i =
\frac{1}{T+1}\sum_{t=0}^{T} \widehat{\bfPhi}_i(\mathbf{p}^t_{-i}). $
Clearly $\mathbf{p}_i^{T+1}$ is an {\it average} of the history of
$S_i$'s water-filling solution, hence the name of Average Iterative
Water-Filling (A-IWF) for the proposed algorithm. This algorithm
maintains the distributed nature of the original IWF, because in
each iteration $t+1$, $S_i$ only needs to know the set of IPN
$\{IPN^t_i(k)\}_{k\in\mathcal{K}}$ as well as its own power
allocation $\{{p}_i^t(k)\}_{k\in\mathcal{K}}$ in iteration $t$ (both
of which can be obtained locally by $S_i$), but {\it does not} need
to know the transmission power profiles of other users.

We see that the main difference between the proposed algorithm and
the previously mentioned R-IWF is that we use a set of {\it
diminishing} and {\it iteration dependent} stepsize
$\{\alpha_t\}_{t=0}^{\infty}$ that satisfies
\eqref{eqAlphaProperty3}, instead of the fixed stepsize $\lambda$.
We will see later that it is exactly these properties of the
$\{\alpha_t\}_{t=0}^{\infty}$ that guarantee the convergence of
A-IWF under IPN uncertainty.

We model the noisy IPN for user $S_i$ on channel $k$ as: $
 \widehat{IPN}^t_i(k)=IPN^t_i(k)+\epsilon^t_i(k)
$, %and the noisy version of the water-filling operator can thus be
%expressed as: $ \widehat{\Phi}_i(\mathbf{p}^t_{-i})=\left[
%{\sigma}_i-{IPN}^t_i(k)-\epsilon^t_i(k) \right]^{p_{mask}(k)}_0$,
where $ \epsilon^t_i(k)$ represents the estimation error of the true
value $IPN^t_i(k)$. Let $\bfepsilon_i\triangleq
[\epsilon_i(1),\cdots,\epsilon_i(K)]^{\intercal}$, and
$\bfepsilon\triangleq
[\bfepsilon_1,\cdots,\bfepsilon_N]^{\intercal}$. Let
$\mathcal{F}_i^T$ be defined as the filtration generated by
$\mathbf{p}_i^{T+1}\bigcup\{\mathbf{p}_i^t,
\widehat{\bfPhi}_i(\mathbf{p}_{-i}^t)\}_{t=0}^{T}$. We assume the
error process to be zero mean, i.e.,
$E[\epsilon^t_i(k)|\mathcal{F}^{t-1}_i]$=0. This assumption is
reasonable because conditioning on the knowledge of the desired
signal ($\mathbf{p}^t_i$ in our case), the estimation error of
$IPN^t_i(k)$, $\epsilon^t_i(k)$ can indeed be viewed as a zero mean
random variable using most conventional estimators (see Section V of
\cite{benedict67} for detailed comparison of estimation biases for
different algorithms). The above model is very general in the sense
that we do not assume the explicit forms of the algorithms that
perform the estimation, nor do we require that the error process
$\{\epsilon^t_i(k)\}_{t=1}^{T}$ be independent with the history of
IPN up to time $T$, i.e., our model allows $\widehat{IPN}^t_i(k)$ to
be calculated based on the previous or the current observations made
by the receiver of $S_i$.

In the following, we use ``w. p. 1" to abbreviate ``with probability
1".   %Assume the noise process
%has the following two properties: A-1)
%$E(\bfepsilon^{t+1}|\mathcal{F}_t)=\mathbf{0}$; A-2) the noise
%process are bounded. %Hence, we have
%%%$E[||\alpha_t \bfepsilon^t||^2|\mathcal{F}^t]\leq (\alpha_t
%%%c^*)^2$, and
%%$\sum_{t=1}^{\infty}||\alpha_t \bfepsilon^t||^2<\infty$ w. p. 1.
We need the following definition before introducing Lemma
\ref{lemmaContraction}, which characterizes the noisy version of the
water-filling operator $\widehat{\bfPhi}(\mathbf{p})$. For any
positive $N\times 1$ vector
$\mathbf{w}\triangleq[w_1,\cdots,w_N]^\intercal$ and the operator
$\widehat{\bfPhi}(\mathbf{p})$, the (vector) {\it block-maximum
norm} $||.||_{2,block}^{\mathbf{w}}$ is defined as
\cite{bertsekas97}: $
||\widehat{\bfPhi}(\mathbf{p})||_{2,block}^{\mathbf{w}}\triangleq
\max_{i\in\mathcal{N}}\frac{||\widehat{\bfPhi}_i(\mathbf{p}_{-i})||_2}{w_i}.
$
\newtheorem{L1}{Lemma}
\begin{L1}\label{lemmaContraction}
{\it Define a $N\times N$ matrix $\bfUpsilon$ related to the channel
gains as:
\begin{align}
[\bfUpsilon]_{i,j}\triangleq \left\{ \begin{array}{ll}
\max_{k\in\mathcal{K}}|\bar{H}_{j,i}(k)|^2 &\textrm{if } i\ne j\\
0  &\textrm{otherwise } \\
\end{array} \right. \label{eqDefineS}.
\end{align}
Let $\rho(\bfUpsilon)$ be the spectral radius of the matrix
$\bfUpsilon$. Then if $\rho(\bfUpsilon)<1$, there must exist a
positive vector $\bar{\mathbf{w}}$, and a constant $\beta$ that
satisfies $0<\beta<1$,
 such that for any feasible
$\mathbf{p}^1,\mathbf{p}^2\in\mathcal{P}$,
\begin{align}
||\widehat{\bfPhi}(\mathbf{p}^1)-{\bfPhi}(\mathbf{p}^2)||_{2,block}^{\bar{\mathbf{w}}}\le
\beta||\mathbf{p}^1-\mathbf{p}^2||_{2,block}^{\bar{\mathbf{w}}}+
||\bfepsilon||_{2,block}^{\bar{\mathbf{w}}}.\label{eqLemma2}
\end{align}}
\end{L1}
\begin{proof}
The Proof is similar to Proposition 2 of \cite{scutari08b}. Please
see Appendix \ref{appendix1} for detail.
\end{proof}

We note here that it has been proven by \cite{scutari08b}, that when
$\rho(\bfUpsilon)<1$ is true, the original water-filling operator
$\bfPhi(\mathbf{p})$ is a contraction with coefficient $\beta<1$,
and hence has a {\it unique} fixed point, i.e., there exists a
unique $\mathbf{p}^*\in\mathcal{P}$ such that
$\mathbf{p}^*=\bfPhi(\mathbf{p}^*)$.

We then characterize the convergence property of the A-IWF algorithm
under two different assumptions of the noise process
$\{\bfepsilon^t\}_{t=0}^{\infty}$. For simplicity of notation, in
the following, we use $||.||$ to denote the norm
$||.||_{2,block}^{\bar{\mathbf{w}}}$, where $\bar{\mathbf{w}}$ is
the positive vector obtained from the proof of Lemma
\ref{lemmaContraction}.

\newtheorem{T1}{Theorem}
\begin{T1}\label{theoremConvergence}
{\it Assume $\rho(\bfUpsilon)<1$, $\{\alpha_t\}_{t=0}^{\infty}$
satisfies \eqref{eqAlphaProperty3}, and
$\{\bfepsilon^t\}_{t=0}^{\infty}$ satisfies $
\sum_{t=1}^{\infty}\alpha_t ||\bfepsilon^t||<\infty,~\textrm{w. p.
1.} $ Then the sequence of power profiles
$\{\mathbf{p}^t\}_{t=1}^{\infty}$ generated by the A-IWF algorithm
converges to the unique fixed point of the original mapping
$\bfPhi(.)$, denoted by $\mathbf{p}^{\ast}$. More precisely, we
have: $ ||\mathbf{p}^t-\mathbf{p}^*||\to 0 \textrm{~w. p. 1}.
$}\end{T1}
\begin{proof}
Please see Appendix \ref{appendixTheorem1} for proof.
\end{proof}

%\newtheorem{R1}{Remark}
%\begin{R1}
%In the above modeling, we summarize all the uncertainties regarding
%to the computation of the water-filling solution as a noise term
%added to the true solution. We can also consider the situation in
%which the IPN is the only uncertainty involved in the computation of
%the water-filling solution. According to classical estimation theory
%\cite{kay93}, we consider a widely used relationship between the
%true value $IPN^t_i(k)$ and its estimates $\widehat{IPN}^t_i(k)$:
%\begin{align}
%\widehat{IPN}^t_i(k)=IPN^t_i(k)+\nu^t_i(k)\label{eqUncertaintyIPN}
%\end{align}
%where $\nu^t_i(k)$ is a zero mean random variable representing
%estimation error, and is independent of the true value $IPN^t_i(k)$.
%Arguing similarly as in the proof of Theorem
%\ref{theoremConvergence}, we can show that the proposed algorithm
%converges almost surely. Due to space limit, we omit the proof.
%\end{R1}

\newtheorem{T2}{Theorem}
\begin{T1}\label{theoremConvergenceAlternative}
{\it Assume  $\rho(\bfUpsilon)<1$, and $\{\alpha_t\}_{t=0}^{\infty}$
satisfies \eqref{eqAlphaProperty3}, and the error process satisfies
$ \lim_{t\to\infty} ||\bfepsilon^t||=0,~\textrm{w. p. 1.} $ Then we
have: $ ||\mathbf{p}^t-\mathbf{p}^*||\to 0 \textrm{~w. p. 1}.
$}\end{T1}
\begin{proof}
Please see Appendix \ref{appendixTheorem2} for proof.
\end{proof}
At this point, we would like to give some remarks regarding to the
above convergence results.
%We point out that these theorems state that the algorithm converges
%``with probability 1", which means that for every realization of the
%algorithm, we have $\lim_{T\to\infty}\mathbf{p}^T=\mathbf{p}^*$
%except for on a set of measure zero.

\newtheorem{R1}{Remark}
\begin{R1}
The condition $\rho(\bfUpsilon)<1$, which is a restriction on the
channel gains, coincides with the condition that ensures the
convergence of IWF without the IPN uncertainties in Theorem 1 of
\cite{scutari08b}. We refer the readers to \cite{scutari08b} for
physical interpretation as well as the comparison of this condition
with other similar conditions derived in the literature, e.g., those
in \cite{luo06b}  and \cite{shum07}.
\end{R1}

%\newtheorem{R2}{Remark}
%\begin{R1}
%In the ideal case when the transmitter can obtain the exact value of
%the IPN, it can be straightforwardly shown that the A-IWF algorithm
%converges whenever the IWF algorithm converges.
%\end{R1}

\newtheorem{R4}{Remark}
\begin{R1}
We will show in section \ref{secSimulation2} that in many cases when
$\rho(\bfUpsilon)<1$ is not satisfied, our algorithm still
converges. This suggests that the A-IWF algorithm may need more
relaxed convergence conditions than the one stated in this
correspondence. We will leave this task as a future research topic.
\end{R1}

Theorem \ref{theoremConvergence} and Theorem
\ref{theoremConvergenceAlternative} differ in their respective
restrictions on the error process $\{\bfepsilon^t\}_{t=0}^{\infty}$,
as technically the conditions $
\sum_{t=0}^{\infty}\alpha_t||\bfepsilon^t||<\infty ~\textrm{and}~
\lim_{t\to\infty}||\bfepsilon^t||=0 $ do not imply each other.
Although these conditions require that the error process be
{\it diminishing}, %Such
%restrictions might be justified by the fact that the receivers are
%getting more and more samples as the iteration of the algorithm
%increases, consequently they are getting better estimation results.
we do observe in our simulations (to be shown in Section
\ref{secSimulation}) that the A-IWF converges in the presence of
more general forms of noises, for example noises with zero mean and
bounded second moment. This observation leads us to believe that the
above conditions on the error process are overly restrictive. Such
belief is partially justified as follows.

Assume
$\hspace{-0.0cm}E[\bfepsilon_i^t|\mathcal{F}_i^{t-1}]\hspace{-0.1cm}=\hspace{-0.1cm}0$,
and $\bfepsilon_i^t$ has bounded second moment for all $i$. Further
assume $\widehat{\bfPhi}(\mathbf{p}^t)$ can be approximated as: $
\widehat{\bfPhi}(\mathbf{p}^t)={\bfPhi}(\mathbf{p}^t)+\bfxi^t $,
where the elements of the {\it bias vector} $\bfxi^t$ satisfies:
\begin{align}
\hspace{-0.0cm}
 E[\xi_i^{t}(k)|\mathcal{F}_i^{t-1}]\hspace{-0.1cm}=\hspace{-0.1cm}{0} \textrm{~and~}
E[(\xi_i^{t}(k))^2|\mathcal{F}_i^{t-1}]\hspace{-0.1cm}<\hspace{-0.1cm}\infty.\label{eqErrorProcess}
\end{align}
Then we have the following convergence result. See Appendix
\ref{appendix2} for the proof.

%\newtheorem{R5}{Remark}
%\begin{R1}
%We also note here that each users $i\in\mathcal{N}$ can choose {\it
%its own} sequence of sepsizes, say
%$\{\alpha_{i,t}\}_{t=1}^{\infty}$, and all the results derived in
%this paper are still valid. Throughout this paper we use the same
%set of sequence $\{\alpha_{t}\}_{t=1}^{\infty}$ for all the users
%for ease of presentation.
%\end{R1}

\newtheorem{T3}{Theorem}
\begin{T1}\label{theoremConvergenceUnbiasedPhi}
{\it Suppose $\widehat{\bfPhi}(\mathbf{p}^t)$ is approximated as $
\widehat{\bfPhi}(\mathbf{p}^t)={\bfPhi}(\mathbf{p}^t)+\bfxi^t$ with
the elements of $\bfxi^t$ satisfies \eqref{eqErrorProcess}. If
$\bfPhi(.)$ is a contraction with constant $\beta$, and if
$\{\alpha_t\}_{t=1}^{\infty}$ satisfy \eqref{eqAlphaProperty3}, then
we have: $ ||\mathbf{p}^t-\mathbf{p}^*||\to 0 \textrm{~w. p. 1}$.}
\end{T1}

Theorem \ref{theoremConvergenceUnbiasedPhi} essentially says that if
the above approximation of the noisy water-filling solution is
accurate, then we only require the error process
$\{\bfepsilon^t\}_{t=0}^{\infty}$ to have mean zero and bounded
second moments to ensure the convergence of the algorithm. Note that
in this case the bias vector $\bfxi^t$ summarizes the uncertainties
regarding both the IPNs and the dual variables. The key assumption
here is that $E[\xi_i^{T}(k)|\mathcal{F}_i^{T-1}]={0}~\forall~i,k$,
i.e., based on all the knowledge it has for the evolution of the
algorithm until time $T-1$, a particular user $S_i$ predicts that
the biases $\{\xi^T_i(k)\}_{k}$ are zero mean. The following
empirical experiments show that such assumption is approximately
true.

\begin{figure*}[htb] \vspace*{-.1cm}
        \begin{minipage}[t]{0.3\linewidth}
    \centering
    {\includegraphics[width=
1\linewidth]{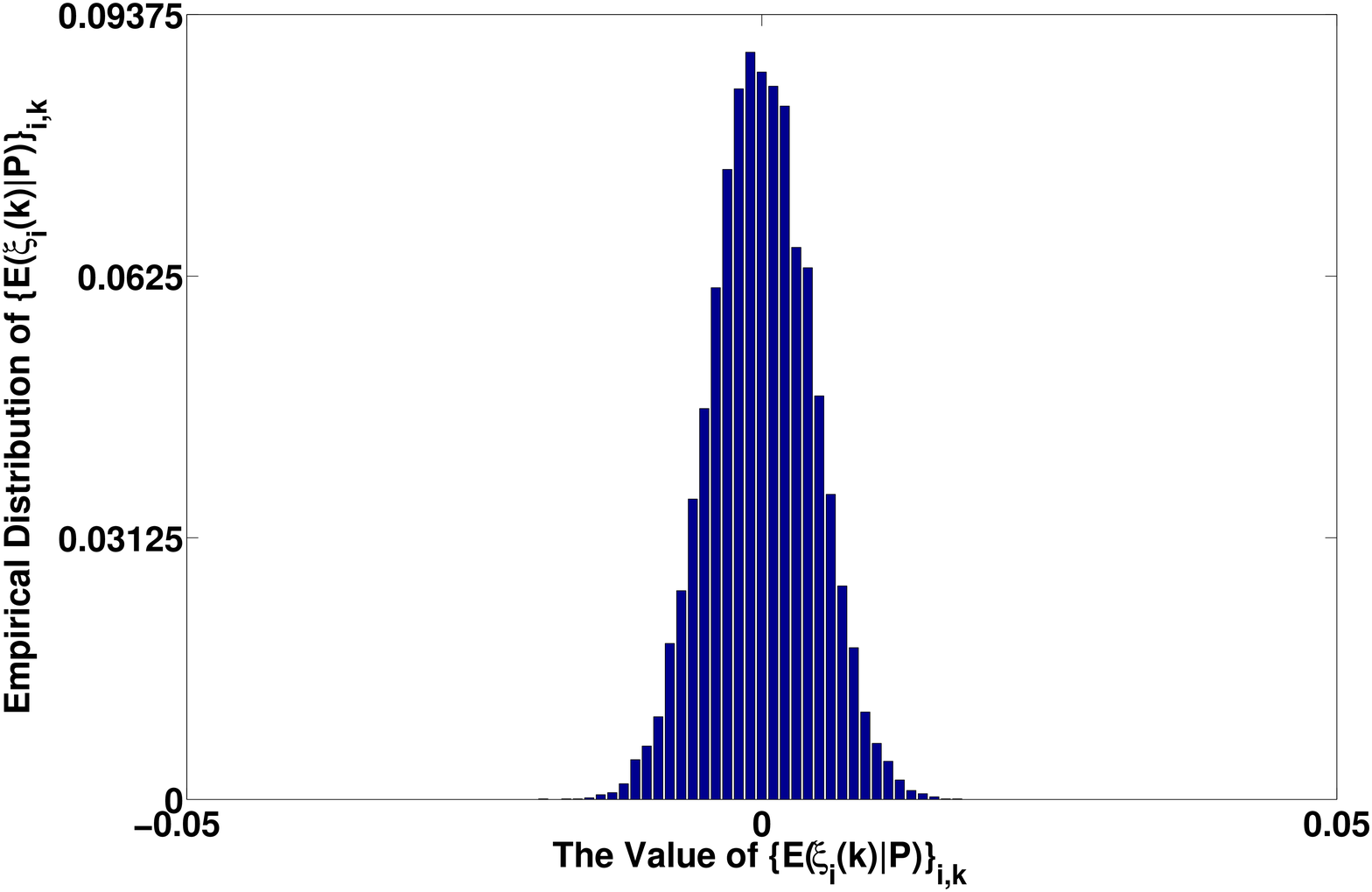} \vspace*{-.1cm}}
\end{minipage}
\begin{minipage}[t]{0.3\linewidth}
    \centering
    {\includegraphics[width=
1\linewidth]{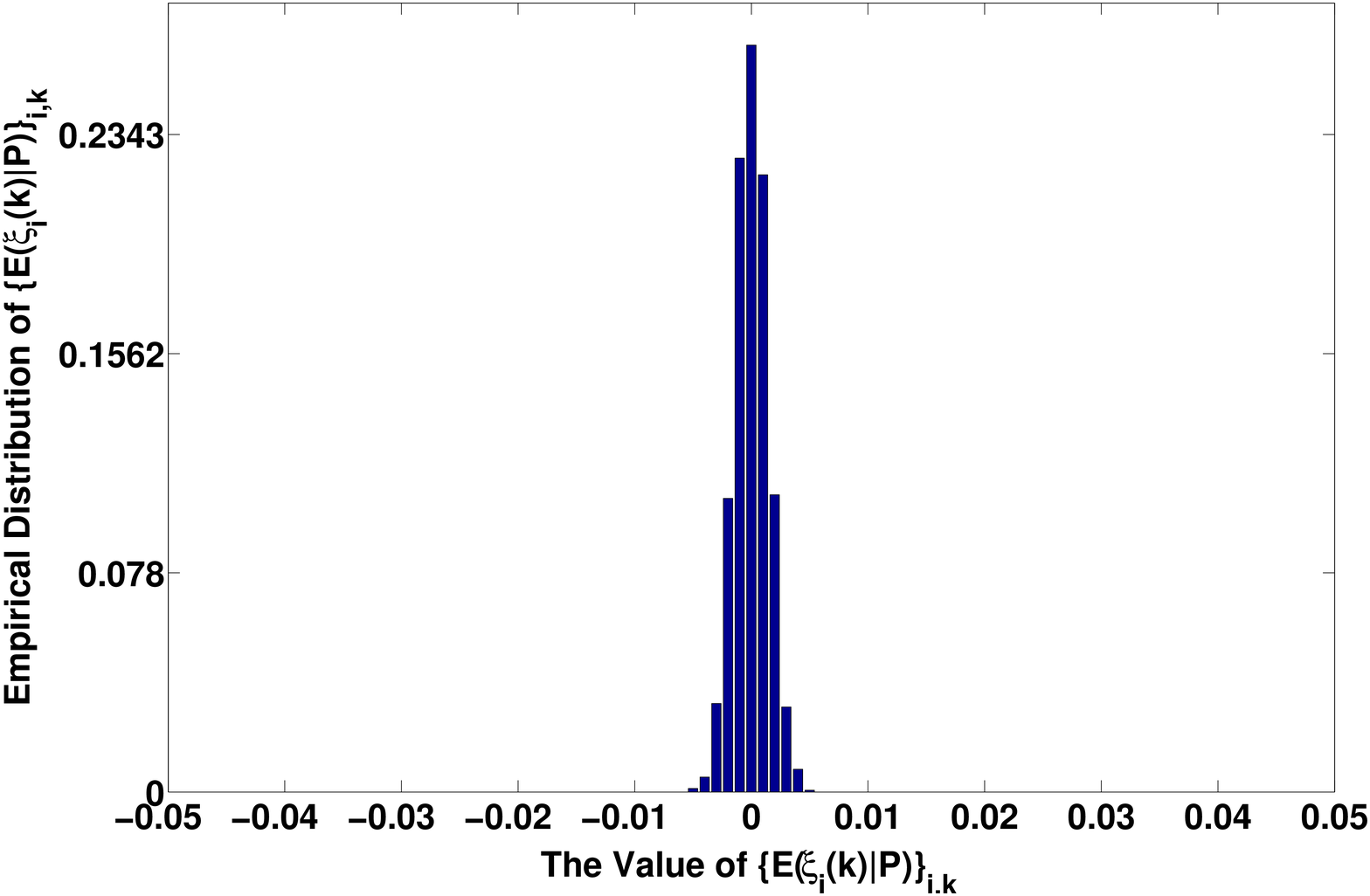} \vspace*{-.1cm}}
\end{minipage}
\begin{minipage}[t]{0.3\linewidth}
    \centering
    {\includegraphics[width=
1\linewidth]{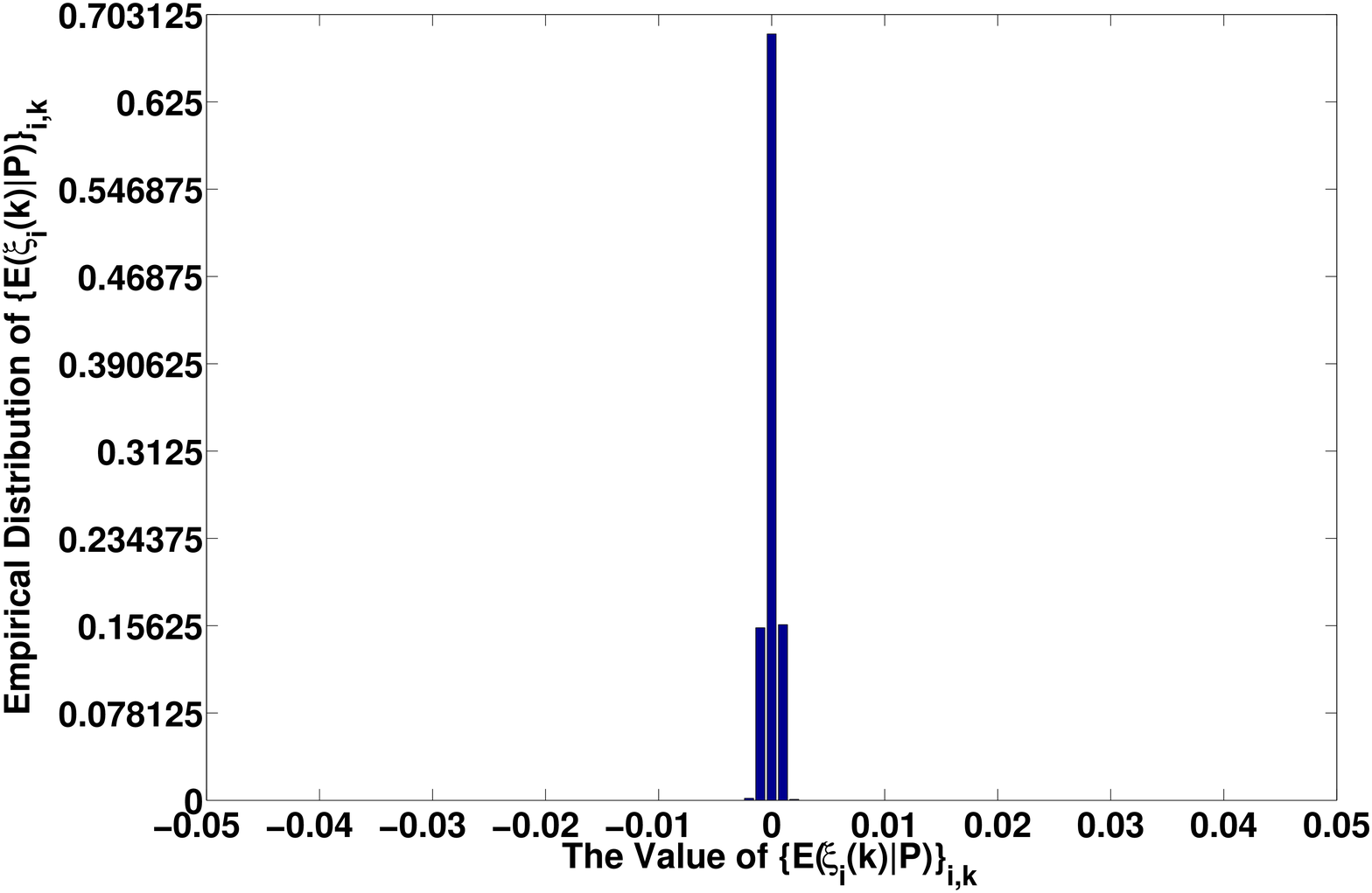} \vspace*{-.1cm}}
\end{minipage}
\vspace*{-0.2cm}\caption{Empirical distribution of
{$\{E[\xi_i(k)|\mathbf{p}]\}_{i,k}$}. Left: $L=1,000$; Middle:
$L=10,000$; Right: $L=100,000$.
}\label{figBiasDistribution}\vspace*{-0.1cm}
    \end{figure*}

Consider a network with 10 users and 32 channels. Let
$\bar{p}_i=10$, $p_{mask}(k)=3,~\forall~k\in\mathcal{K}$.  %We
%compare the difference between the noisy water-filling solution
%calculated by:
%$\widehat{\Phi}^k_i(\mathbf{p}_{-i})=\left[\widehat{\sigma}_i-IPN^t_i(k)-\epsilon_i(k)\right]^{p_{mask}(k)}_0$
%and the exact water-filling solution:
%${\Phi}^k_i(\mathbf{p}_{-i})=\left[{\sigma}_i-IPN^t_i(k)\right]^{p_{mask}(k)}_0$
%when $\mathbf{p}$ is known.
We define the bias of the noisy water-filling solution as:
\begin{align}
\xi_i(k)&\triangleq\widehat{\Phi}^k_i(\mathbf{p}_{-i})-{\Phi}^k_i(\mathbf{p}_{-i})\nonumber\\
&=\left[\widehat{\sigma}_i-IPN_i(k)-\epsilon_i(k)\right]^{p_{mask}(k)}_0-\left[{\sigma}_i-IPN_i(k)\right]^{p_{mask}(k)}_0.\nonumber
\end{align}
We simplify the analysis a bit by assuming the bias process to be
Markovian, i.e.,
$E[\xi_i^{T}(k)|\mathcal{F}_i^{T-1}]=E[\xi_i^{T}(k)|\mathbf{p}_i^{T}]$.
We investigate the distribution of
$\{E[\xi_i(k)|\mathbf{p}_i]\}_{i,k}$. Define the variance of noise
$\epsilon_i(k)$ as $var_i(k)$; introduce a term called Interference
Error Ratio (IER) to quantify the strength of the IPN error
$\bfepsilon$: $ IER_i(k)\triangleq
10\log_{10}\left(\frac{IPN_i(k)}{var_i(k)}\right)$. We fix $IER=10
dB$ during the experiment. As $E[\xi_i(k)|\mathbf{p}_i]$ is a
function of $\mathbf{p}_i$, we fix
$\{\mathbf{p}_i\in\mathcal{P}_i\}_{i\in\mathcal{N}}$, and obtain an
estimate of $\{E[\xi_i(k)|\mathbf{p}_i]\}_{i,k}$, denoted by
$\{{M}_i(k)\}_{i,k}$, by doing the follows: 1) generate the channel
gains $\{|H_{i,j}(k)|^2)\}$ randomly; 2) generate $L$ samples of IPN
noise vectors $\{\bfepsilon^{l}\}_{l=1}^{L}$ by:
$\epsilon_i^{l}(k)\sim N(0, var_i(k)),~\forall~i,k,y$; 3) obtain the
bias $\{\bfxi^{l}\}_{l=1}^{L}$ according to its definition above; 4)
calculate ${M}_i(k)=
\frac{1}{L}\sum_{l=1}^{L}\xi^{l}_i(k),~\forall~i,k$. We repeat the
above procedure for 1,000 times with randomly generated sets of
$\{\mathbf{p}_i\in\mathcal{P}_i\}_{i\in\mathcal{N}}$, and plot the
empirical distribution of $\{E[\xi_i(k)|\mathbf{p}_i]\}_{i,k}$ in
Fig. \ref{figBiasDistribution} (different graphs in Fig.
\ref{figBiasDistribution} represent the results obtained by
experiments using different $L$). We see that when the estimates
$\{{M}_i(k)\}_{i,k}$ are getting more accurate with larger number of
samples (larger $L$), the empirical distribution of
$\{E[\xi_i(k)|\mathbf{p}_i]\}_{i,k}$ is getting more concentrated at
zero. Thus we conjecture that asymptotically with $L\to\infty$,
$E[\xi_i(k)|\mathbf{p}_i]$ can be approximated as zero for all $i$
and $k$.

%Let the superscript $(m)$ index each run of the experiment.
%
%In $m^{th}$ run of the experiment, we do the following: 1) generate
%$\mathbf{p}^{(m)}_i\in\mathcal{P}_i,~ \forall~i\in\mathcal{N}$
%randomly and keep them fixed during this run; 2) generate $Y$
%samples of IPN noise vectors $\{\bfepsilon^{y,(m)}\}_{y=1}^{Y}$,
%where the elements of $\bfepsilon^{y,(m)}$ is generated by normal
%distributions: $\epsilon_i^{y,(m)}(k)\sim N(0, var_i^{(m)}(k))$; 3)
%calculate the bias $\{\bfxi^{y,(m)}\}_{y=1}^{Y}$. We call
%${M}^{(m)}_i(k)\triangleq \frac{1}{Y}\sum_{n=1}^{Y}\xi^{n,(m)}_i(k)$
%the $m^{th}$ sample of the empirical mean of the bias $\xi_i(k)$. We
%conduct 1000 runs of the above experiment, and plot the histogram of
%$\{{M}^{(m)}_i(k)\}_{i, k, m}$ in the left part of Fig.
%\ref{figBiasDistribution}.

%The above results show that in a particular iteration, when
%$\mathbf{p}$ is known but the estimated IPN is unknown due to the
%IPN noises, the empirical mean of the biases of the noisy
%water-filling solutions centers around zero. Particularly, the
%percentage of the samples of empirical mean of the bias that locate
%within the interval of $(-0.01,0.01)$ are
%$\%97.6,~\%99.8,~\textrm{and}~ \%100$ respectively for these three
%experiments. The conclusion draw from the this empirical study is
%that $E[\xi_i^{T}(k)|\mathcal{F}_{T-1}]$ is approximated around
%zero, thus the convergence results in Theorem
%\ref{theoremConvergenceUnbiasedPhi} can be viewed as approximately
%true.

\newtheorem{R5}{Remark}
\begin{R1}
We give some remarks comparing the convergence conditions of
conventional IWF and A-IWF under uncertainty. From \cite{ortega72}
(Chapter 12, Th. 12.2.1--12.2.5) we see that the condition $
\lim_{t\to\infty} ||\bfepsilon^t||=0$ in Theorem
\ref{theoremConvergenceAlternative} is sufficient and necessary for
the conventional IWF to converge to the fixed point without
performing averaging. However, the conventional IWF diverges under
condition $ \sum_{t=1}^{\infty}\alpha_t ||\bfepsilon^t||<\infty$ in
Theorem \ref{theoremConvergence}, because this condition is not
equivalent to $ \lim_{t\to\infty} ||\bfepsilon^t||=0$. Moreover,
from Th. 12.2.5 in \cite{ortega72}, under the assumption in Theorem
\ref{theoremConvergenceUnbiasedPhi}, the conventional IWF produces a
sequence that finally stays in a ball around the fixed point.
However, the radius of such ball is increasing with
$\frac{\max_t{||\bfxi^t||}}{1-\beta}$. Notice that in this case
$||\bfxi^t||$ needs not to be decreasing, thus the maximum possible
error of the conventional IWF may be large (consider when $\beta$ is
close to $1$).
\end{R1}

\section{Simulation Results}\label{secSimulation}
In this section we conduct three experiments to demonstrate the
properties of the A-IWF algorithm.
%In
%section \ref{secSimulation1}, we show that the proposed algorithm is
%robust against estimation error compared with IWF algorithm. In
%section \ref{secSimulation2}, we show the proposed algorithm
%converges in scenarios where $\bfPhi(.)$ is not a contraction, in
%which situation the IWF algorithm diverges. Finally, we compare the
%proposed algorithm with the IWF when the we assume that the
%water-filling solution can be computed \emph{accurately}, and the
%mapping $\bfPhi(.)$ is a contraction. The purpose of this simulation
%is to show that the A-IWF converges fast in ideal scenarios.
\vspace*{-.3cm}
\subsection{Performance with Estimation Error}\label{secSimulation1}
We simulate a network with $10$ randomly located users, and $64$
channels. We choose the noise to be a zero mean Gaussian random
variable as $\epsilon^t_i(k)\sim N(0,var^t_i(k))$; we choose the
$IER$ for all the users on all the channels to be
$IER^t_{i}(k)=20dB,~15dB$; we choose the channel gains
$\{|H_{i,j}(k)|^2\}$ randomly and appropriately such that
$\rho(\bfUpsilon)<1$ is satisfied; we choose
$\alpha_t=\frac{1}{t+1}$. For ease of demonstration, different
algorithms are examined with the same starting points.

\begin{figure*}[ht] \vspace*{0cm}\centerline{\includegraphics*[width=
0.5\linewidth]{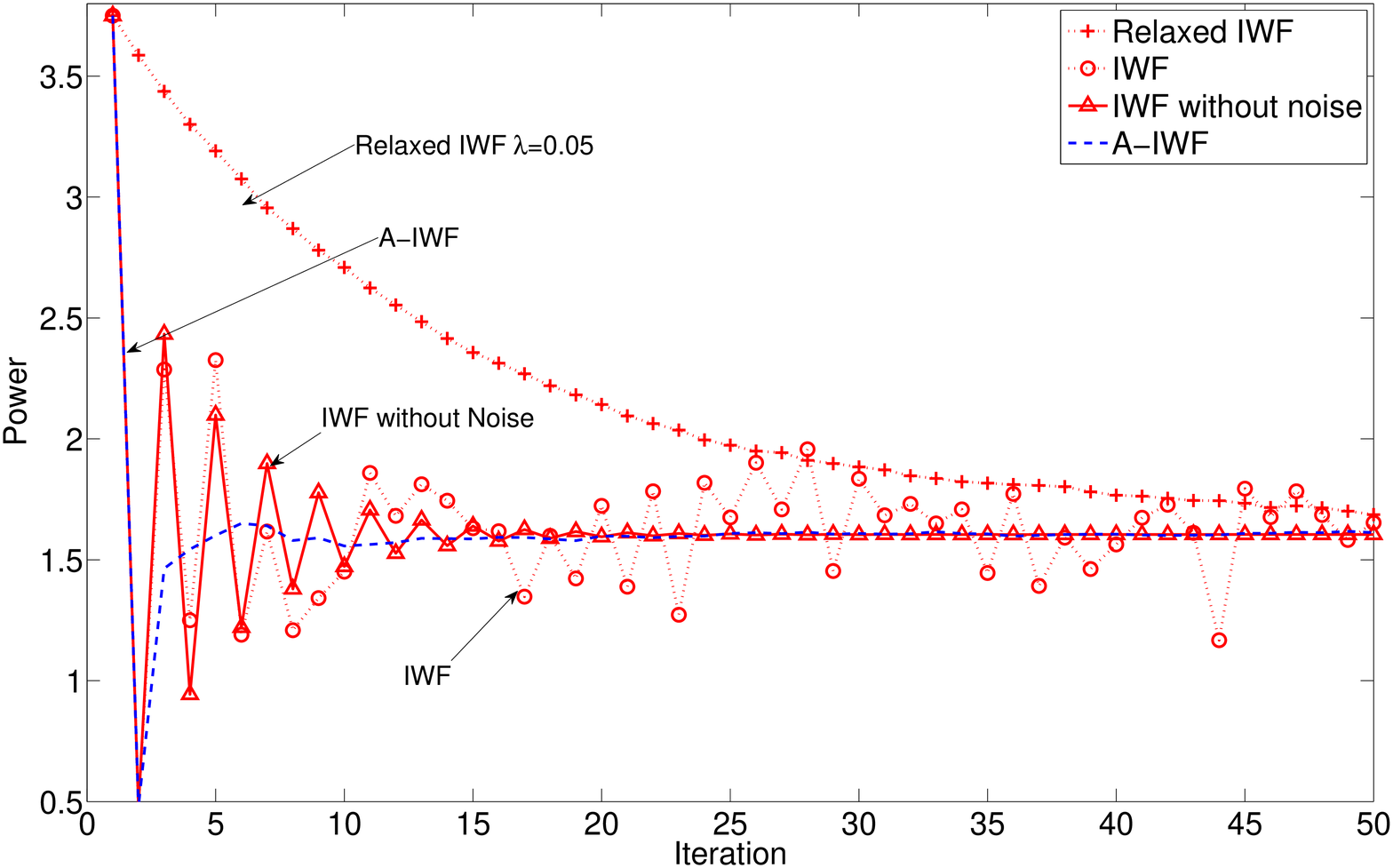}
\includegraphics*[width=
0.5\linewidth]{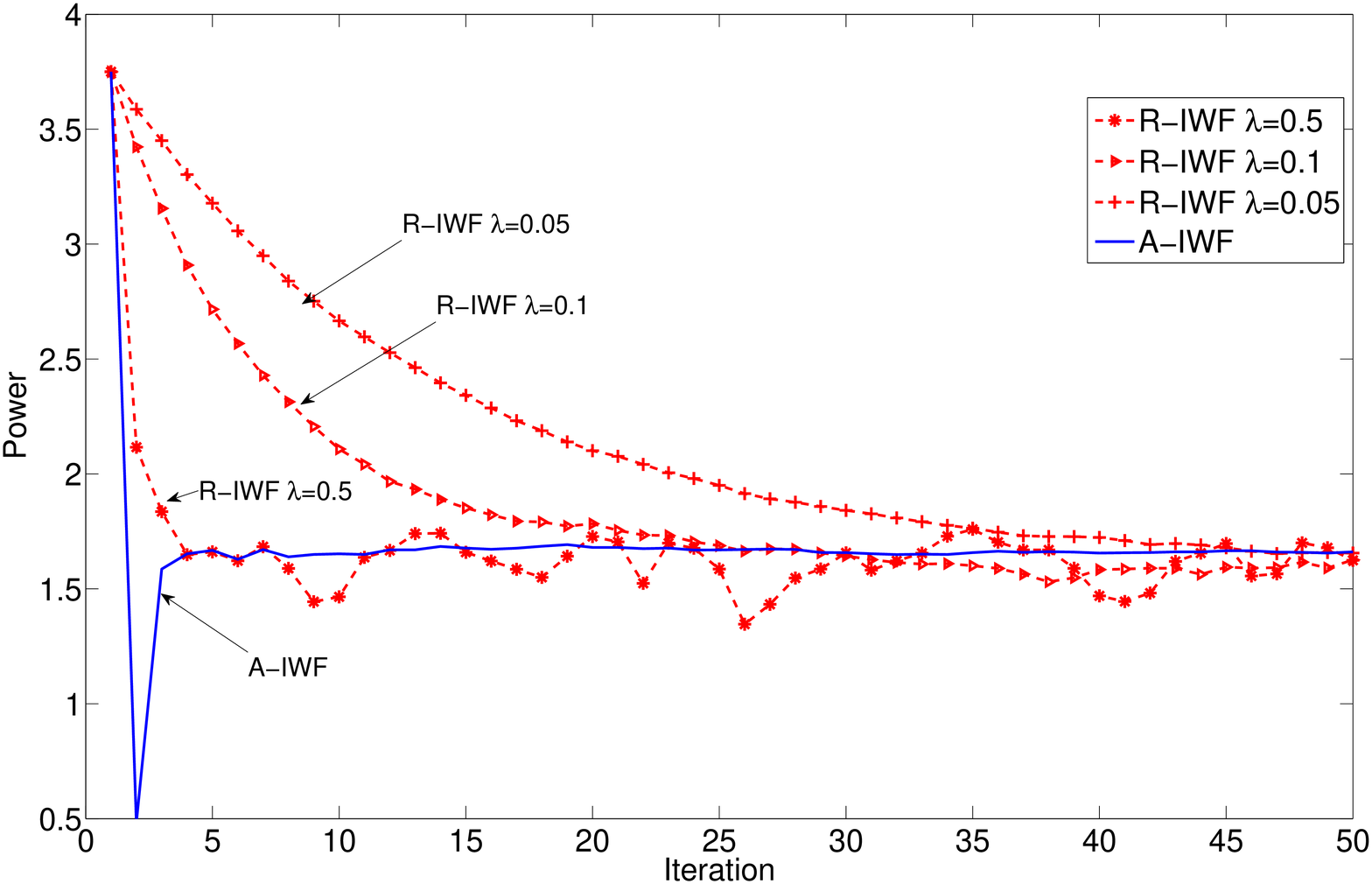}\vspace*{-.3cm}}
    \caption{Comparison of the output for different
    algorithms, IER=20dB.}\label{figNoisyIWF20}    \vspace*{-.2cm}
    \end{figure*}

\begin{figure*}[ht] \vspace*{0cm}
    \begin{minipage}[t]{0.5\linewidth}
    \centering
    {\includegraphics[width=
1\linewidth]{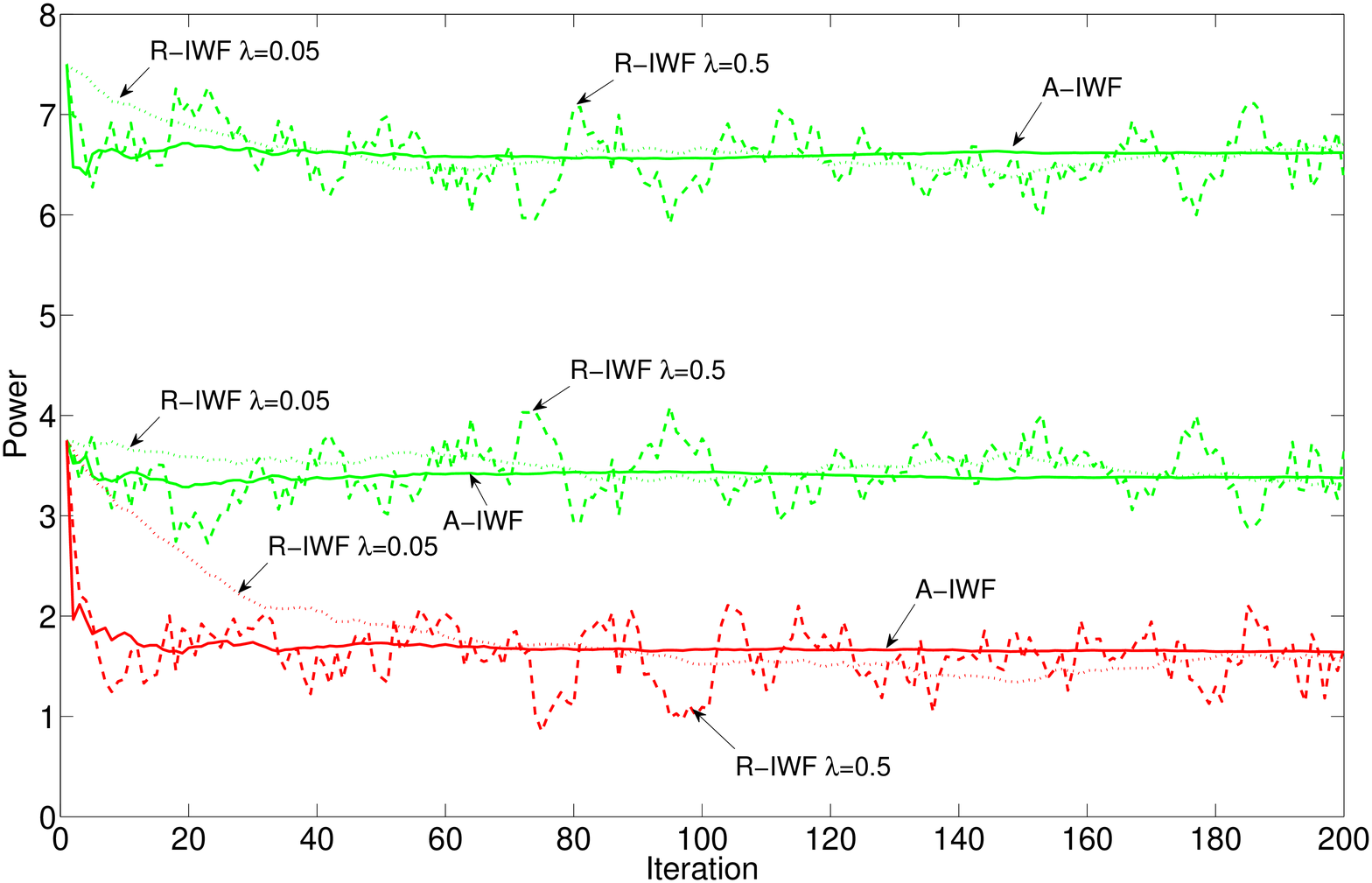} \vspace*{-0.2cm}\caption{Comparison of
different algorithms, IER=15dB.}\label{figNoisyIWF15}
\vspace*{-0.2cm}}
\end{minipage}
    \begin{minipage}[t]{0.5\linewidth}
    \centering
    {\includegraphics[width=
1\linewidth]{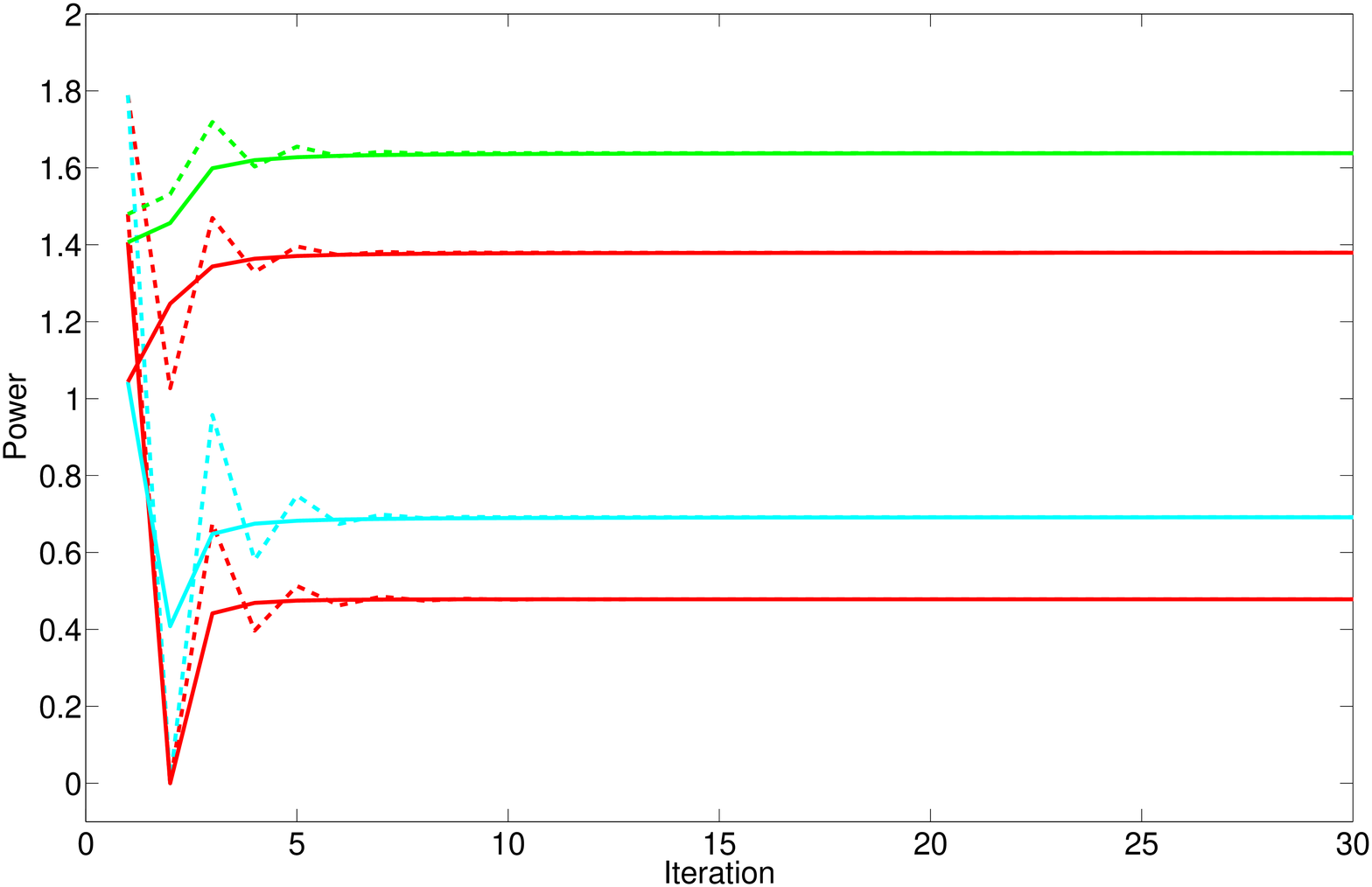}
\vspace*{-0.2cm}\caption{Comparison of convergence speed of IWF and
A-IWF.}\label{figSpeed} \vspace*{-0.2cm}}
\end{minipage}
    \end{figure*}

In Fig. \ref{figNoisyIWF20}, we show the power output produced by
various algorithms of a particular user on a particular channel,
with $IER=20dB$. It is clear that in the presence of estimation
error, the IWF algorithm produces a sequence of noisy power profiles
which exhibits no sign of convergence. We also show the performance
of IWF algorithm \emph{without} estimation error, for the purpose of
comparison. It is seen that the A-IWF algorithm converges to the
unique NE predicted by the IWF (without estimation error) quickly.
In  Fig. \ref{figNoisyIWF20}, we also show the output of the R-IWF
algorithm with various values of $\lambda$. We observe that when
$\lambda$ is large, the output is still noisy, while when $\lambda$
is small, the convergence is slow. The point is that the choice of
$\lambda$ is important for the performance of R-IWF, but it is
difficult to correctly choose $\lambda$ to guarantee both robustness
and fast convergence. In Fig. \ref{figNoisyIWF15}, we compare the
selected power profiles of R-IWF and A-IWF when $IER=15dB$.

\subsection{Performance with Strong Interference}\label{secSimulation2}
As stated above, the convergence of the IWF in ideal situations
usually dependes on the weak interference condition. It is observed
that in the system with strong interference, IWF algorithm diverges
\cite{leshem09}. In the following simulation, we demonstrate several
scenarios in which the IWF diverges, but the A-IWF algorithm
converges. The purpose of these simulations is to argue that the
A-IWF may need weaker conditions for convergence.

\begin{figure*}[ht] \vspace*{-.2cm}\centerline{\includegraphics*[width=
0.45\linewidth]{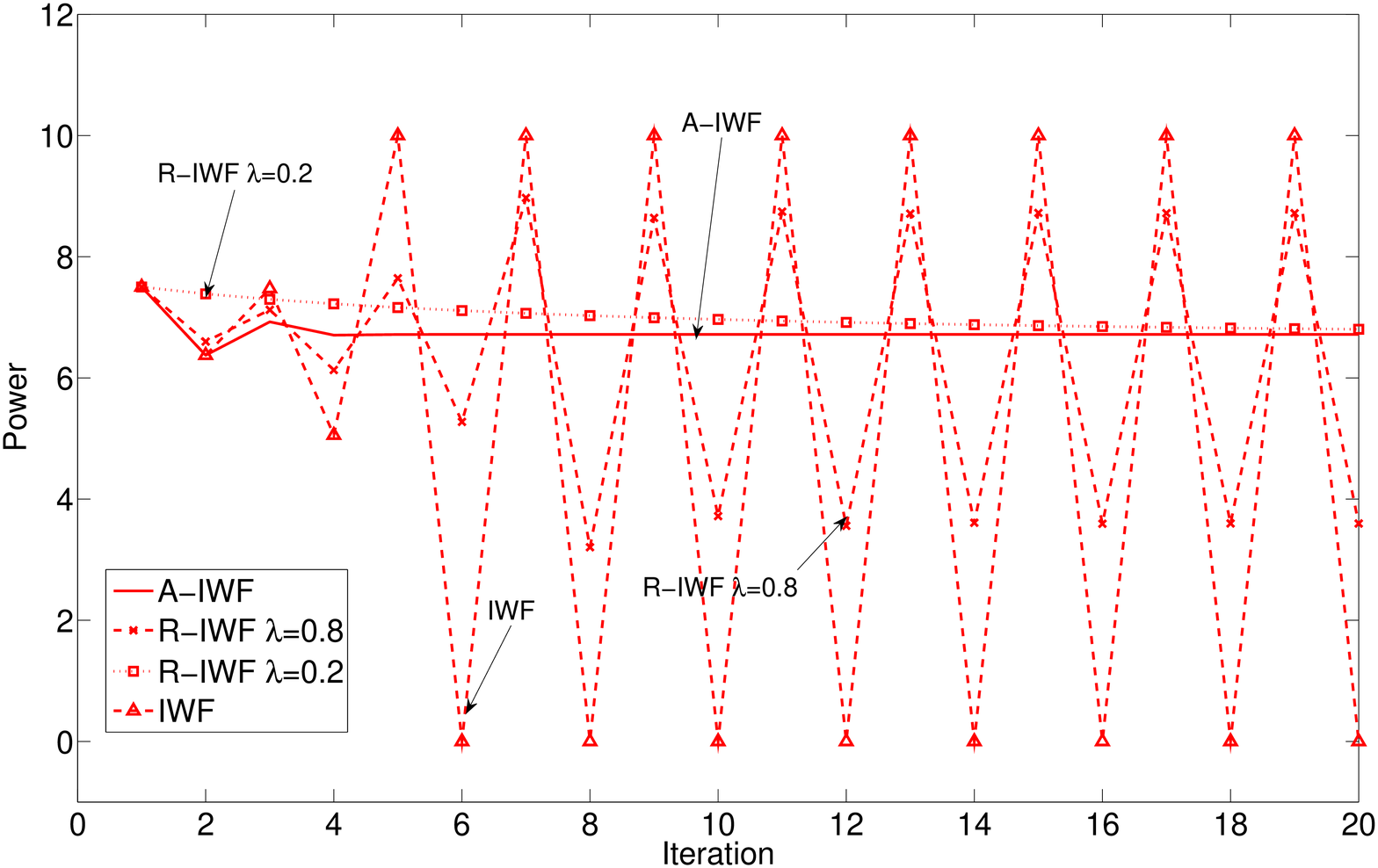}\includegraphics*[width=
0.45\linewidth]{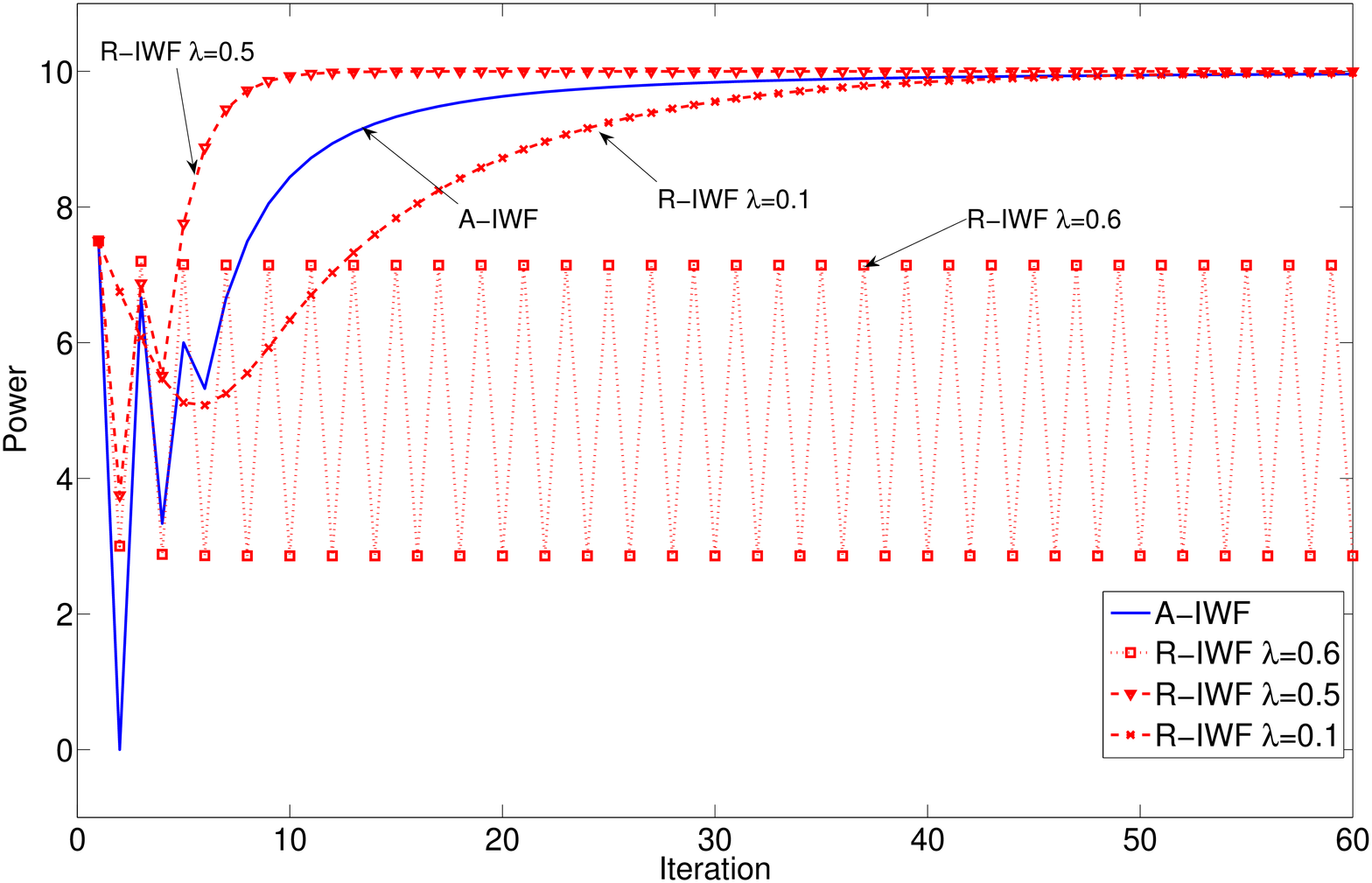} }\vspace*{-.3cm}
    \caption{Convergence properties of difference algorithms in strong interference channels.}\label{figIWFDiverge1}
 \vspace*{-.2cm}
    \end{figure*}

Consider the following scenario of strong interference (example $5$
in \cite{leshem09}). Suppose there are $3$ users and $2$ channels in
the system, with channel matrices $\mathbf{H}(k)$ expressed as
follows:
\begin{align}
{\tiny \mathbf{H}(1)= \mathbf{H}(2)=\left(\begin{array}{ccc}
1&0&2\\
2&1&0\\
0&2&1
\end{array}\right)}\nonumber
\end{align}
where each element of the matrix $\mathbf{H}(k)$ is defined as
$[\mathbf{H}(k)]_{i,j}\triangleq |H_{i,j}(k)|^2$. Set the noise
power on channel $1$ to be $\sigma^2$, the noise on channel $2$ set
to be $\sigma^2+\bar{p}_i$, with $\bar{p}_i=10$, for all
$i\in\mathcal{N}$. There is a unique NE of this game, in which each
user allocates two-thirds of its power to channel $1$ and the rest
to channel $2$. The left hand side part of Fig. \ref{figIWFDiverge1}
shows the power profiles of user $1$ on channel $1$ that are
produced by different algorithms (with the same starting point). It
is seen that the IWF algorithm oscillates, while the A-IWF algorithm
converges quickly. Similar results are obtained in the right hand
side part of Fig. \ref{figIWFDiverge1} with the following settings:
\begin{align}
\tiny \mathbf{H}(1)= \left(\begin{array}{ccc}
1&2&4\\
4&1&2\\
2&4&1
\end{array}\right),~
\mathbf{H}(2)= \left(\begin{array}{ccc}
2&3&5\\
3&2&5\\
5&3&2
\end{array}\right)
\end{align}
and the noise power on both channels set to be the same. We observe
again that the performance of R-IWF algorithm is very sensitive to
the choice of $\lambda$: when $0.6\le \lambda \le 1$, the output
oscillates; when $0\le \lambda\le 0.5$, the output converges, with
larger $\lambda$ for faster convergence. However, it is not clear
what rules one should follow in general to select such critical
parameter.

%    \begin{figure}[ht] {\includegraphics[width=
%0.5\linewidth]{figIWFDiverge2.eps} }
%    \caption{Comparison of the output for different
%    algorithms.}\label{figIWFDiverge2}
%    \end{figure}

\subsection{Convergence In Ideal Cases}\label{secSimulation3}
%In this subsection, we demonstrate that in ideal cases, the A-IWF
%algorithm converges as fast as IWF algorithm.
Questions may arise as to how does the A-IWF perform in situations
when the water-filling solution in \eqref{eqWaterFilling} can be
carried out accurately. As shown in \cite{scutari08b}, the IWF
algorithm converges \emph{linearly} in this ideal scenario.
Theoretically, we can only show that A-IWF converges {\it
sublinearly} in ideal scenario, i.e.,
$\lim_{t\to\infty}\frac{||\mathbf{p}^{t+1}-\mathbf{p}^*||}{||\mathbf{p}^t-\mathbf{p}^*||}=1$.
%However, in simulation it seems that the A-IWF algorithm converge as
%fast as IWF.%The simulation is still based on the $10$ user $64$ channel network,
%and we assume perfect estimation of the IPN and weak interference.
However, we observe in various randomly generated channel gains and
random starting points of the algorithms, that the A-IWF algorithm
seems to always converge as fast as the IWF algorithm. Fig.
\ref{figSpeed} shows such an instance of this experiment. In this
figure, we compare the power output of selected users on selected
channels (in a network with 10 users and 64 channels) generated by
the IWF and the A-IWF. It takes less than $10$ iterations before two
algorithms agree with each other. Note that the dotted lines
represent the output of the IWF algorithms and the
 solid lines represent the output of the A-IWF algorithm.
%    \begin{figure}[ht] \vspace*{-.5cm}{\includegraphics[width=
%0.5\linewidth]{figCompareIWFSpeed.eps}  \vspace*{-.3cm}}
%    \caption{Comparison of convergence speed of the IWF and the A-IWF.
%    \newline
%    The dotted lines represent the output of the IWF algorithms; the
%    solid lines represent the output of the A-IWF algorithm}\label{figSpeed} \vspace*{-.7cm}
%    \end{figure}

\section{Conclusion}\label{conclusion}
In this correspondence, we proposed an extension to the IWF
algorithm which is more robust and has better convergence
properties. We proved that the proposed algorithm converges w. p. 1
under suitable assumptions. We argue that this algorithm is indeed
robust against time-varying estimation error of the power of
interference plus noise that is needed for the computation of the
IWF computation. We also show by simulation that the proposed
algorithm converges when strong interferences are present in the
communication channel, a scenario in which the IWF algorithm
diverges. An interesting future research topic is to develop a
possibly more general condition for the convergence of the proposed
algorithm.

 \appendices

\section{Proof of Lemma
\ref{lemmaContraction}}\label{appendix1}
\begin{proof}
Define {
$\bar{\mathbf{H}}_{j,i}=diag(|\bar{H}_{j,i}(1)|^2,\cdots,|\bar{H}_{j,i}(K)|^2)$};
define {
$\mathbf{IPN}_i=\left[IPN_i(1),\cdots,IPN_i(K)\right]^{\intercal}$},
and define $\widehat{\mathbf{IPN}}_i$ similarly. From Corollary 3 of
\cite{scutari08b}, we have that the water-filling operator {
$\bfPhi_i(\mathbf{p}_{-i})$} can be expressed as the projection of
$-\mathbf{IPN}_i$ onto the space $\mathcal{P}_i$, i.e., {  $
\bfPhi_i(\mathbf{p}_{-i})=\left[-\mathbf{IPN}_i\right]_{\mathcal{P}_i}.
$} Similarly, we have that {
$\widehat{\bfPhi}_i(\mathbf{p}_{-i})=\left[-\widehat{\mathbf{IPN}}_i\right]_{\
\mathcal{P}_i} $}. Consequently, we have:
\begin{align}
&||\widehat{\bfPhi}_i(\mathbf{p}^1_{-i})-{\bfPhi}_i(\mathbf{p}^2_{-i})||_2\stackrel{(a)}
\le||-\sum_{j\ne
i}\bar{\mathbf{H}}_{j,i}\mathbf{p}^1_j-\bfepsilon_i+\sum_{j\ne
i}\bar{\mathbf{H}}_{j,i}\mathbf{p}^2_j||_2\nonumber\\
& \le\sum_{j\ne
i}||\bar{\mathbf{H}}_{j,i}||_2||\mathbf{p}^1_j-\mathbf{p}^2_j||_2+||\bfepsilon_i||_2\nonumber\\
&\stackrel{(b)}=\sum_{j\ne
i}\left(\max_k|\bar{{H}}_{j,i}(k)|^2\right)||\mathbf{p}^1_j-\mathbf{p}^2_j||_2+||\bfepsilon_i||_2,~\forall~i\in\mathcal{N}\label{eq2NormContraction}
\end{align}
where $(a)$ is because of the non-expansiveness of the projection
operator under Euclidean norm; $(b)$ is due to the fact that the
2-norm of a diagonal matrix equals to the maximum absolute value of
its diagonal elements. Define
${e}_{\bfPhi_i}\triangleq||\widehat{\bfPhi}_i(\mathbf{p}^1_{-i})-{\bfPhi}_i(\mathbf{p}^2_{-i})||_2$,
${e}_i\triangleq ||\mathbf{p}^1_i-\mathbf{p}^2_i||_2$, and let
$\mathbf{e}_{\bfPhi}\triangleq\left[{e}_{\bfPhi_1},\cdots,{e}_{\bfPhi_N}\right]^{\intercal}$,
$\mathbf{e}\triangleq\left[{e}_{1},\cdots,{e}_{N}\right]^{\intercal}$,
and
$\mathbf{e}_{\bfepsilon}\triangleq\left[||\bfepsilon_1||_2,\cdots,||\bfepsilon_N||_2\right]^{\intercal}$.

In order to proceed, we define the vector weighted maximum norm
\cite{bertsekas97} as:
\begin{align}
||\mathbf{x}||^{\mathbf{w}}_{\infty, vec}\triangleq
\max_{i}\frac{|x_i|}{w_i}, ~\mathbf{w}>\mathbf{0},
\mathbf{x}\in\mathbb{R}^{N}
\end{align}
and the matrix weighted maximum norm as:
\begin{align}
||\mathbf{A}||^{\mathbf{w}}_{\infty, mat}\triangleq
\max_{i}\frac{1}{w_i}\sum_{j=1}^{N}{|[\mathbf{A}]_{i,j}|}w_j,
~\mathbf{w}>\mathbf{0}, \mathbf{A}\in\mathbb{R}^{N\times N}.
\end{align}
Notice, that from the definition of norm
$||.||^{\mathbf{w}}_{\infty, vec}$, $||.||^{\mathbf{w}}_{\infty,
mat}$ and the block-maximum norm, we have the following equivalence:
\begin{align}
||\mathbf{e}||^{\mathbf{w}}_{\infty,
vec}&=\max_{i}\frac{|{e}_i|}{w_i}
=\max_{i}\frac{||\mathbf{p}_i^1-\mathbf{p}_i^2||}{w_i}=||\mathbf{p}^1-\mathbf{p}^2||^{\mathbf{w}}_{2, block}\nonumber\\
||\mathbf{e}_{\bfPhi}||^{\mathbf{w}}_{\infty,
vec}&=\max_{i}\frac{|e_{\bfPhi_i}|}{w_i}=||\widehat{\bfPhi}(\mathbf{p}^1)-\bfPhi(\mathbf{p}^2)||^{\mathbf{w}}_{2,
block}\nonumber\\
 ||\mathbf{e}_{\bfepsilon}||^{\mathbf{w}}_{\infty,
vec}&=\max_{i}\frac{||\bfepsilon_i||_2}{w_i}
=||\bfepsilon||^{\mathbf{w}}_{2, block} \label{eqNormEquivalence}.
\end{align}
The set of $N$ inequalities in \eqref{eq2NormContraction} can be
concisely written in vector form as ($\bfUpsilon$ is defined in
\eqref{eqDefineS}): $\mathbf{e}_{\bfPhi}\le \bfUpsilon
\mathbf{e}+\mathbf{e}_{\bfepsilon}.$ Applying vector weighted
maximum norm to this inequality results in:
\begin{align}
||\mathbf{e}_{\bfPhi}||^{\mathbf{w}}_{\infty, vec}&\le
||\bfUpsilon||^{\mathbf{w}}_{\infty, mat}
||\mathbf{e}||^{\mathbf{w}}_{\infty,
vec}+||\mathbf{e}_{\bfepsilon}||^{\mathbf{w}}_{\infty,
vec}\nonumber\\
&=||\bfUpsilon||^{\mathbf{w}}_{\infty, mat}
||\mathbf{p}^1-\mathbf{p}^2||^{\mathbf{w}}_{2,
block}+||{\bfepsilon}||^{\mathbf{w}}_{2, block}\label{eqNormMatVec}.
\end{align}

Arguing similarly as the derivation of the Proposition 2 of
\cite{scutari08b} by using \eqref{eqNormEquivalence} and
\eqref{eqNormMatVec} we have:
\begin{align}
&||\widehat{\bfPhi}(\mathbf{p}^1)-\bfPhi(\mathbf{p}^2)||^{\mathbf{w}}_{2,
block}\nonumber\\&=||\mathbf{e}_{\bfPhi}||^{\mathbf{w}}_{\infty,
vec} \le||\bfUpsilon||^{\mathbf{w}}_{\infty,
mat}||\mathbf{p}^1-\mathbf{p}^2||_{2,block}^{{\mathbf{w}}}+
||\bfepsilon||_{2,block}^{{\mathbf{w}}}.
\end{align}
Since $\bfUpsilon$ is a non-negative matrix, from \cite{bertsekas97}
Corollary 6.1, we have that there exists a $\bar{{\mathbf{w}}}$ such
that $ \rho(\bfUpsilon)<1 \Longleftrightarrow
||\bfUpsilon||_{\infty,mat}^{\bar{{\mathbf{w}}}}<1. $ Consequently,
we conclude that if $\rho(\bfUpsilon)<1 $, then there must exists a
$\beta\in(0,1)$ and a positive vector $\bar{{\mathbf{w}}}$ that
satisfy \eqref{eqLemma2}.
\end{proof}

\section{Proof of Theorem
\ref{theoremConvergence}}\label{appendixTheorem1}
\begin{proof}
Starting from an arbitrary initial point
$\mathbf{p}^0\in\mathcal{P}$, the magnitude of the difference
between $\mathbf{p}^1$ and the fixed point $\mathbf{p}^{\ast}$ can
be expressed as:
\begin{align}
||\mathbf{p}^{1}-\mathbf{p}^*||&=||(1-\alpha_0)\mathbf{p}^0+\alpha_0\widehat{\bfPhi}(\mathbf{p}^0)-\bfp^{*}||\nonumber\\
&\le ||(1-\alpha_0)(\mathbf{p}^0-\mathbf{p}^*)||
+\left|\left|\alpha_0\left(\widehat{\bfPhi}(\mathbf{p}^0)-{\bfPhi}(\mathbf{p}^*)\right)\right|\right|\nonumber\\
&\stackrel{(i)}\leq
(1-\alpha_0)||\mathbf{p}^{0}-\mathbf{p}^*||+\alpha_0\beta||\mathbf{p}^{0}-\mathbf{p}^*||+||\alpha_0\bfepsilon^0|| \nonumber\\
%&=\left(1-\alpha_0(1-\beta)\right)||\bfp^0-\bfp^*||+||\alpha_0\bfepsilon^0||\nonumber\\
&=\left(1-\alpha_0(1-\beta)+\frac{||\alpha_0\bfepsilon^0||}{||\bfp^0-\bfp^*||}\right)||\bfp^0-\bfp^*||\label{eqInductionBase}
\end{align}
where $(i)$ is from Lemma \ref{lemmaContraction}. Let us denote
$\mu_t\triangleq(1-\alpha_t(1-\beta))$. From \eqref{eqAIWF} and
\eqref{eqAlphaProperty3}, clearly we have $\alpha_0=1$ and
$\alpha_t\le 1$,
which implies $\mu_0\le \mu_t~\forall~t$. %Then the
%second iteration is as follows: \vspace{-0.2cm}{
%\begin{align}
%||\bfp^2-\bfp^*||&\leq
%(1-\alpha_1)||\bfp^1-\bfp^*||+\alpha_1\beta||\bfp^1-\bfp^*||+||\alpha_1\bfepsilon^1||\nonumber\\
%&=(1-\alpha_1(1-\beta))||\bfp^1-\bfp^*||+||\alpha_1\bfepsilon^1||=\mu_1\left(\mu_0+\frac{||\alpha_0\bfepsilon^0||}{||\bfp^0-\bfp^*||}\right)||\bfp^0-\bfp^*||+||\alpha_1\bfepsilon^1||\nonumber\\
%&=\left(\mu_0\mu_1+\mu_1\frac{||\alpha_0\bfepsilon^0||}{||\bfp^0-\bfp^*||}+\frac{||\alpha_1\bfepsilon^1||}{||\bfp^0-\bfp^*||}\right)||\bfp^0-\bfp^*||.
%\end{align}}
By induction, we show that in general:
\begin{align}
&||\bfp^T-\bfp^*||\leq \Big(\prod_{t=0}^{T-1}\mu_i+
\frac{\sum_{t=0}^{T-1}\left(\prod_{j=t}^{T-1}\mu_j\right)||\alpha_t\bfepsilon^t||}{\mu_0||\bfp^0-\bfp^*||}\Big)||\bfp^0-\bfp^*||
\label{eqDifferenceP}.
\end{align}
Clearly from \eqref{eqInductionBase} at time $T=1$,
\eqref{eqDifferenceP} is true. Suppose at time $T$,
\eqref{eqDifferenceP} is true. At time $T+1$, we have:
\begin{align}
&||\bfp^{T+1}-\bfp^*||\leq
(1-\alpha_T)||\bfp^T-\bfp^*||+\alpha_T\beta||\bfp^T-\bfp^*||+||\alpha_T\bfepsilon^T||\nonumber\\
&\le\Big(\prod_{t=0}^{T}\mu_i+
\frac{\sum_{t=0}^{T-1}\left(\prod_{j=t}^{T}\mu_j\right)||\alpha_t\bfepsilon^t||}{\mu_0||\bfp^0-\bfp^*||}+\frac{||\alpha_T\bfepsilon^T||}{||\bfp^0-\bfp^*||}\Big)||\bfp^0-\bfp^*||\nonumber\\
&\le\Big(\prod_{t=0}^{T}\mu_i+
\frac{\sum_{t=0}^{T}\left(\prod_{j=t}^{T}\mu_j\right)||\alpha_t\bfepsilon^t||}{\mu_0||\bfp^0-\bfp^*||}\Big)||\bfp^0-\bfp^*||.
\end{align}
Note that in the last inequality, we have used the fact that
$\mu_t\ge \mu_0$, and
$\frac{||\alpha_T\bfepsilon^T||}{||\bfp^0-\bfp^*||}\le
\frac{\mu_T||\alpha_T\bfepsilon^T||}{\mu_0||\bfp^0-\bfp^*||}$.
%One way to do this is to choose
%$\alpha_t=\frac{\frac{1}{t+2}}{1-\beta}$, which implies that
%\begin{align}
%\lim_{T\to\infty}\prod_{t=0}^{T}\mu_t=
%\lim_{T\to\infty}\frac{1}{2}\times\frac{2}{3}\times\cdots\frac{T+1}{T+2}=0\nonumber.
%\end{align}
From the assumption $\sum_{t=1}^{\infty}\alpha_t
||\bfepsilon^t||<\infty,~\textrm{w. p. 1}$, there must exist some
constant $0<b<\infty$ such that:
\begin{eqnarray}
\lim_{T\to\infty}\sum_{t=1}^{T}||\alpha_t \bfepsilon^t||\le b<
\infty ~\textrm{w. p. 1}\label{eqConvergeSumEpsilon}.
\end{eqnarray}
In the following, we show $
\lim_{T\to\infty}\sum_{t=0}^{T-1}\left(\prod_{j=t}^{T-1}\mu_j\right)||\alpha_t\bfepsilon^t||=0~
$ w. p. 1.

First note that we have $\lim_{T\to\infty}\prod_{t=0}^{T}\mu_t=0$,
because:
\begin{align}
\lim_{T\to\infty}\log\Big(\prod_{t=0}^{T}\mu_t\Big)
&=\lim_{T\to\infty}\sum_{t=0}^{T}\log\Big(1+(-\alpha_t(1-\beta))\Big)\stackrel{(i)}\nonumber\\
&\le
\lim_{T\to\infty}(1-\beta)\sum_{t=0}^{T}-\alpha_t\stackrel{(ii)}=-\infty\label{eqProductMu}
\end{align}
where $(i)$ is because {  $-1<-\alpha_t(1-\beta)$} and the fact
$\log(1+x)\le x, \forall~x>-1$, $(ii)$ is because
\eqref{eqAlphaProperty3} and $\beta<1$. Clearly \eqref{eqProductMu}
implies {  $\lim_{T\to\infty}\prod_{t=0}^{T}\mu_t=0$}. Thus for any
$\delta>0$, and a fixed $T$ there exists $\hat{T}(T,\delta)>T$ such
that:
\begin{eqnarray}
\prod_{t=T}^{N-1}\mu_t \leq \frac{\delta}{2 b},~\forall~N\geq
\hat{T}(T,\delta).\label{eqConvergeProductMuDelta}
\end{eqnarray}
From \eqref{eqConvergeSumEpsilon} we have that for any $\delta>0$,
there exists $T(\delta)$ such that:
\begin{eqnarray}
\sum_{t=T}^{\infty}||\alpha_t\bfepsilon^t||\leq
\frac{\delta}{2},~\forall~T\geq T(\delta),~\textrm{w. p.
1}.\label{eqConvergeSumEpsilonDelta}
\end{eqnarray}
Then we have that for all
$N>\max\left\{T(\delta),\hat{T}(T(\delta),\delta)\right\}=\hat{T}(T(\delta),\delta)$:
\begin{align}
&\sum_{t=0}^{N}\big(\prod_{j=t}^{N-1}\mu_j\big)||\alpha_t\bfepsilon^t||\nonumber\\
&=\sum_{t=0}^{T(\delta)}\big(\prod_{j=t}^{N-1}\mu_j\big)||\alpha_t\bfepsilon^t||+
\sum_{t=T(\delta)+1}^{N}\big(\prod_{j=t}^{N-1}\mu_j\big)||\alpha_t\bfepsilon^t||\nonumber\\
&\stackrel{(i)}\leq
\sum_{t=0}^{T(\delta)}\big(\prod_{j=T(\delta)}^{N-1}\mu_j\big)||\alpha_t\bfepsilon^t||+\frac{\delta}{2}\nonumber\\
&\stackrel{(ii)}\leq
\prod_{j=T(\delta)}^{N-1}\mu_j\sum_{t=0}^{T(\delta)}||\alpha_t\bfepsilon^t||+\frac{\delta}{2}\stackrel{(iii)}\nonumber\\
&\leq \frac{\delta}{2 b}b+\frac{\delta}{2}=\delta~~\textrm{w. p. 1}
\end{align}
where $(i)$ is because \eqref{eqConvergeSumEpsilonDelta} and the
fact that $\prod_{j=t}^{N-1}\mu_j<1$ for all $t\leq N-1$; $(ii)$ is
because $\prod_{j=T(\delta)}^{N-1}\mu_j$ is independent of $t$;
$(iii)$ is because of \eqref{eqConvergeSumEpsilon} and
\eqref{eqConvergeProductMuDelta}. Consequently, we have that:
\begin{eqnarray}
\lim_{T\to\infty}\sum_{t=0}^{T-1}\big(\prod_{j=t}^{T-1}\mu_j\big)||\alpha_t\bfepsilon^t||=0~
\textrm{w. p. 1}\label{eqConvergeSumProduct}.
\end{eqnarray}
From \eqref{eqDifferenceP}, \eqref{eqConvergeSumProduct}, and
$\lim_{T\to\infty}\prod_{t=0}^{T}\mu_t=0$, we conclude : $
\lim_{t\to\infty}||\bfp^t-\bfp^*||=0~\textrm{w. p. 1}. $
%Additionally, if there is no IPN error, \eqref{eqConvergeSumEpsilon}
%is automatically true, and the algorithm also converges.
\end{proof}

\section{Proof of Theorem
\ref{theoremConvergenceAlternative}}\label{appendixTheorem2}
\begin{proof}
Due to space limit, we only show the proof for the case that
$\alpha_t=\frac{1}{t+1}$. The proof for general $\{\alpha_t\}$ can
be obtained similarly. When taking $\alpha_t=\frac{1}{t+1}$, the
A-IWF algorithm can be written compactly as:
$\mathbf{p}^{T+1}=\frac{1}{T+1}\sum_{t=0}^{T}\widehat{\bfPhi}(\mathbf{p}^t)$.
We can write:
\begin{align}
||\mathbf{p}^{T+1}-\mathbf{p}^*||&\le
\frac{1}{T+1}\sum_{t=0}^{T}||\widehat{\bfPhi}(\mathbf{p}^t)-{\bfPhi}(\mathbf{p}^*)||\nonumber\\
&\stackrel{(i)}\le
\frac{\beta}{T+1}\sum_{t=0}^{T}||\mathbf{p}^t-\mathbf{p}^*||+\frac{1}{T+1}\sum_{t=0}^{T}||\bfepsilon^t||\label{eqAverageDifference}
\end{align}
where $(i)$ is from Lemma \ref{lemmaContraction}. Suppose the
sequence $\{\mathbf{p}^t\}$ does not converge to $\mathbf{p}^*$,
i.e., $\lim_{T\to\infty}\sup||\mathbf{p}^T-\mathbf{p}^*||=\delta>0$.
Using the Stolz-Ces\`aro Theorem \cite{muresan08}, we have that:
\begin{align}
\hspace{-0.4cm}&\lim_{T\to\infty}\sup\frac{\sum_{t=0}^{T}||\mathbf{p}^t-\mathbf{p}^{\ast}||}{T+1}\le
\lim_{T\to\infty}\sup||\mathbf{p}^T-\mathbf{p}^{\ast}||=\delta;
\nonumber\\
&\lim_{T\to\infty}\frac{\sum_{t=0}^{T}||\bfepsilon^t||}{T+1}=\hspace{-0.1cm}\lim_{T\to\infty}||\bfepsilon^T||=0,~\textrm{w.
p. 1}.\label{eqCesaro}
\end{align}
Taking $\lim\sup$ on both sides of \eqref{eqAverageDifference}, we
have:
\begin{align}
&\lim_{T\to\infty}\sup||\mathbf{p}^{T+1}-\mathbf{p}^*||\nonumber\\
&\le
\lim_{T\to\infty}\sup\frac{\beta}{T+1}\sum_{t=0}^{T}||\mathbf{p}^t-\mathbf{p}^*||+\lim_{T\to\infty}\sup\frac{1}{T+1}\sum_{t=0}^{T}||\bfepsilon^t||
\end{align}
which can be reduced to: $ \delta\le\beta\delta$ by applying
\eqref{eqCesaro}. This is a contradiction to the fact that
$\beta<1$. Then we conclude that
$\lim_{T\to\infty}\sup||\mathbf{p}^T-\mathbf{p}^*||=0$ which in turn
implies $\lim_{T\to\infty}||\mathbf{p}^T-\mathbf{p}^*||=0$.
\end{proof}

\section{Proof of Theorem
\ref{theoremConvergenceUnbiasedPhi}}\label{appendix2} Due to space
limit, we only show the proof for the case that
$\alpha_t=\frac{1}{t+1}$. The proof for general $\{\alpha_t\}$ can
be obtained similarly. We first state a lemma, the proof of which
can be found in Appendix \ref{appendix3}.
\newtheorem{L4}{Lemma}
\begin{L1}\label{lemmaW}
{\it If ${w}^{t+1}=(1-\alpha_t){w}^{t}+\alpha_t\xi^{t+1}$, and
$E[\xi^t|\mathcal{F}^{t-1}]=0$, and
$E[(\xi^t)^2|\mathcal{F}^{t-1}]=b$ is uniformly bounded,
$\{\alpha_t\}$ satisfies \eqref{eqAlphaProperty3}, then we must have
$ \lim_{t\to\infty}{w}^t=0,~\textrm{w. p. 1.} $}
\end{L1}

We are now ready to prove Theorem
\ref{theoremConvergenceUnbiasedPhi}. The A-IWF algorithm can be
compactly written as: $
\mathbf{p}^{T+1}\hspace{-0.2cm}=\frac{1}{T+1}\sum_{t=0}^{T}\widehat{\bfPhi}(\mathbf{p}^t)\hspace{-0.1cm}
=\hspace{-0.1cm}\frac{1}{T+1}\sum_{t=0}^{T}{\bfPhi}(\mathbf{p}^t)
+\mathbf{w}^T$, where $
\mathbf{w}^T\triangleq\frac{1}{T+1}\sum_{t=0}^{T}\bfxi^t=
(1-\frac{1}{T+1})\mathbf{w}^{T-1}+\frac{1}{T+1}\bfxi^T $. Note that
by applying the results of Lemma \ref{lemmaW}, we have
$\lim_{T\to\infty}\mathbf{w}^T=0$. Then the magnitude of difference
between $\mathbf{p}^{T+1}$ and the unique fixed point of the mapping
$\bfPhi(.)$ can be expressed as:
\begin{align}
||\mathbf{p}^{T+1}-\mathbf{p}^{\ast}||&\le\frac{1}{T+1}\sum_{t=0}^{T}||{\bfPhi}(\mathbf{p}^t)-{\bfPhi}(\mathbf{p}^{\ast})||
+||\mathbf{w}^T||\nonumber\\
&\le
\frac{\beta}{T+1}\sum_{t=0}^{T}||\mathbf{p}^t-\mathbf{p}^{\ast}||
+||\mathbf{w}^T||\label{eqAverageDifference2}.
\end{align}
Suppose the sequence $\{\mathbf{p}^t\}$ does not converge to
$\mathbf{p}^*$, then there must exist a $\delta>0$ such that
$\lim_{T\to\infty}\sup||\mathbf{p}^t-\mathbf{p}^*||=\delta$. Using
again the Stolz-Ces\`aro Theorem as in \eqref{eqCesaro}, and taking
$\lim\sup$ on both sides of \eqref{eqAverageDifference2}, we have: $
\lim_{T\to\infty}\sup||\mathbf{p}^{T+1}-\mathbf{p}^{\ast}||\le
\lim_{T\to\infty}\sup\frac{\beta}{T+1}\sum_{t=0}^{T}||\mathbf{p}^t-\mathbf{p}^{\ast}||
+\lim_{T\to\infty}||\mathbf{w}^T||. $ This inequality can be reduced
to: $\delta\le\beta \delta$, which contradicts to the fact that
$\beta<1$. Thus we conclude that
$\lim_{T\to\infty}||\mathbf{p}^t-\mathbf{p}^*||=0$, and that
$\lim_{T\to\infty}\mathbf{p}^t=\mathbf{p}^{\ast}$.

\section{Proof of Lemma \ref{lemmaW}}\label{appendix3}
\begin{proof}
We have ${w}^{t+1}={w}^t+\alpha_t(\xi^{t+1}-{w}^t)$. Consider the
following iteration:
\begin{align}
({w}^{t+1})^2&=({w}^t+\alpha_t(\epsilon^{t+1}-{w}^t))^2\nonumber\\
&=({w}^t)^2+2\alpha_t(\xi^{t+1}-{w}^t){w}^t+\alpha^2_t(\xi^{t+1}-{w}^t)^2.
\end{align}
Then $E[({w}^{t+1})^2|\mathcal{F}^t]$ can be expressed as:
\begin{align}
&E[({w}^{t+1})^2|\mathcal{F}^t]\nonumber\\
&=({w}^t)^2_t-2\alpha_t
({w}^t)^2+2\alpha_t E[\xi^{t+1}|\mathcal{F}^t]{w}^t+
\alpha^2_t E[(\xi^{t+1}-{w}^t)^2|\mathcal{F}^t]\nonumber\\
&=({w}^t)^2-2\alpha_t
({w}^t)^2+\alpha_t^2\left(E[(\xi^{t+1})^2|\mathcal{F}^t]+({w}^t)^2-2{w}^t
E[\xi^{t+1}|\mathcal{F}^t]\right)\nonumber\\
&\le ({w}^t)^2-2\alpha_t(1-\frac{\alpha_t}{2})({w}^t)^2+\alpha^2_t
b.
\end{align}
Notice that the term $2\alpha_t(1-\frac{\alpha_t}{2})({w}^t)^2\ge 0$
because $0<\alpha_t\le1$. We see that
$\lim_{T\to\infty}\sum_{t=0}^{T}\alpha^2_t b<\infty$ because
$\sum_{t=1}^{\infty}\alpha^2_t<\infty$. In order to proceed, we
define the notion of a {\it non-negative almost-supermartingale}
\cite{robbins71}. Let $z_t$, $\beta_t$, $\xi_t$ and $\zeta_t$ be
{\it non-negative} $\mathcal{F}^t$ measurable random variables. The
sequence $\{z_t\}$ is called {\it non-negative
almost-supermartingale} if $
E[z_{t+1}|\mathcal{F}^t]\le(1+\beta_t)z_t+\xi_t-\zeta_t.$ From
Theorem 1 of \cite{robbins71}, we have $\lim_{t\to\infty}z_t$ exists
and is finite and $\sum_{t=1}^{\infty}\zeta_t<\infty$ w. p. 1 if $
\left\{\sum_{t=1}^{\infty}\beta_t<\infty,
\sum_{t=1}^{\infty}\xi_t<\infty \right\} $.

Now it is clear that the sequence $\{({w}^t)^2\}_{t=0}^{\infty}$ is
a non-negative almost-supermartingale, and according to the above
mentioned theorem we have the following results: 1)
$\{({w}^t)^2\}_{t=0}^{\infty}$ converges; 2)
$\sum_{t=1}^{T}\alpha_t(1-\frac{\alpha_t}{2})({w}^t)^2<\infty$ w. p.
1. The second result implies that $
\lim_{T\to\infty}\sum_{t=1}^{T}\alpha_t ({w}^t)^2<\infty $. Combined
with the fact that $\sum_{t=0}^{\infty}\alpha_t=\infty$ and
$\lim_{t\to\infty}\alpha_t=0$, we have that
$\lim\inf_{t\to\infty}({w}^t)^2={0}$. Moreover, we know from the
first result that the sequence $\{({w}^t)^2\}_{t=0}^{\infty}$
converges, then it must converge to ${0}$.
\end{proof}

\bibliographystyle{IEEEbib}

\bibliography{ref}

\begin{thebibliography}{10}

\bibitem{yu02}
W.~Yu, G.~Ginis, and J.~M. Cioffi,
\newblock ``Distributed multiuser power control for digital subscriber lines,''
\newblock {\em IEEE Journal on Selected Areas in Communications}, vol. 20, no.
  5, pp. 1105--1115, 2002.

\bibitem{scutari08a}
G.~Scutari, D.~P. Palomar, and S.~Barbarossa,
\newblock ``Optimal linear precoding strategies for wideband noncooperative
  systems based on game theory -- part {I}: Nash equilibria,''
\newblock {\em IEEE Trans. on Signal Processing}, vol. 56, no. 3, pp.
  1230--1249, 2008.

\bibitem{wang08}
F.~Wang, M.~Krunz, and S.~G. Cui,
\newblock ``Price-based spectrum management in cognitive radio networks,''
\newblock {\em IEEE Journal of Selected Topics in Signal Processing}, vol. 2,
  no. 1, pp. 74--87, 2008.

\bibitem{scutari08b}
G.~Scutari, D.~P. Palomar, and S.~Barbarossa,
\newblock ``Optimal linear precoding strategies for wideband noncooperative
  systems based on game theory -- part {II}: Algorithms,''
\newblock {\em IEEE Trans. on Signal Processing}, vol. 56, no. 3, pp.
  1250--1267, 2008.

\bibitem{scutari08c}
G.~Scutari, D.~P. Palomar, and S.~Barbarossa,
\newblock ``Asynchronous iterative water-filling for {G}aussian
  frequency-selective interference channels,''
\newblock {\em IEEE Transactions on Information Theory}, vol. 54, no. 7, pp.
  2868--2878, 2008.

\bibitem{luo06b}
Z-.~Q. Luo and J-.S. Pang,
\newblock ``Analysis of iterative waterfilling algorithm for multiuser power
  contorl in digital subscriber lines,''
\newblock {\em EURASIP Journal on Applied Signal Processing}, vol. 2006, pp.
  1--10, 2006.

\bibitem{shum07}
K.~W. Shum, K.~K. Leung, and C.~W. Sung,
\newblock ``Convergence of iterative waterfilling algorithm for {G}aussian
  interference channels,''
\newblock {\em IEEE Journal on Selected Area in Communications}, vol. 25, pp.
  1091--1100, 2007.

\bibitem{leshem09}
A.~Leshem and E.~Zehavi,
\newblock ``Game theory and the frequency selective interference channel,''
\newblock {\em IEEE Signal Processing Magazine}, vol. 26, no. 5, pp. 28--40,
  2009.

\bibitem{benedict67}
T.~R. Benedict and T.~T. Soong,
\newblock ``The joint estimation of signal and noise from the sum evelope,''
\newblock {\em IEEE Transactions on Information Theory}, vol. 13, no. 3, pp.
  447--454, 1967.

\bibitem{pauluzzi00}
D.~R. Pauluzzi and N.~C. Beaulieu,
\newblock ``A comparison of {SNR} estimation techniques for the {AWGN}
  channel,''
\newblock {\em IEEE Transactions on Communications}, vol. 48, no. 10, pp.
  1681--1691, 2000.

\bibitem{setoodeh09}
P.~Setoodeh and S.~Haykin,
\newblock ``Robust transmit powercontrol for cognitive radio,''
\newblock {\em Proceedings of IEEE}, pp. 915--939, 2009.

\bibitem{gohary09b}
R.~H. Gohary and T.~J. Willink,
\newblock ``Robust {IWFA} for open-spectrum communications,''
\newblock {\em IEEE Trans. On Signal Process.}, vol. 57, no. 12, pp.
  4964--4970, 2009.

\bibitem{cheng10}
Y.~Cheng and V.~K.~N. Lau,
\newblock ``Distributive power control algorithm for multicarrier interference
  network over time-varying fading channels--tracking performance analysis and
  optimization,''
\newblock {\em IEEE Transactions on Signal Processing}, vol. 58, no. 9, pp.
  4750--4760, 2010.

\bibitem{mann53}
W.~R. Mann,
\newblock ``Mean value methods in iteration,''
\newblock in {\em Proc. Amer. Math.Soc.}, 1953, pp. 506--510.

\bibitem{bertsekas97}
D.~P. Bertsekas and J.~N. Tsitsiklis,
\newblock {\em Parallel and Distributed Computation: Numerical Methods},
\newblock Athena Scientific, 1997.

\bibitem{ortega72}
J.~M. Ortega and W.~C. Rheinboldt,
\newblock {\em Iterative Solution of Nonlinear Equations in Several Variables},
\newblock Academic Press, 1972.

\bibitem{muresan08}
M.~Muresan,
\newblock {\em A Concrete Approach to Classical Analysis},
\newblock Springer, 2008.

\bibitem{robbins71}
H.~Robbins and D.~Siegmund,
\newblock {\em A Convergence Theorem for Non-Negative Almost Supermartingales
  and Some Applications},
\newblock Optimizing Methods in Statistics. Academic Press, New York, 1971.

\end{thebibliography}

\end{document}